\documentclass[10pt]{article}

\usepackage{amsmath}
\usepackage{amssymb}

\usepackage{graphicx}

\usepackage{cite}

\usepackage{color} 

\usepackage[labelfont=bf,labelsep=period,justification=raggedright]{caption}

\makeatletter
\renewcommand{\@biblabel}[1]{\quad#1.}
\makeatother

\newcommand{\diff}[2]{\frac{\partial #1}{\partial #2}}
\newcommand{\dd}{\,\mathrm{d}}

\date{}

\begin{document}

\begin{flushleft}
{\Large
\textbf{The syncytial \textit{Drosophila} embryo as a mechanically excitable medium}
}
\\
Timon Idema$^{1,\ast}$, 
Julien O. Dubuis$^{2}$,
Louis Kang$^{1}$,
M. Lisa Manning$^{3}$,
Philip C. Nelson$^{1}$,
Tom C. Lubensky$^{1}$,
Andrea J. Liu$^{1,\dag}$
\\
\bf{1} Department of Physics and Astronomy, University of Pennsylvania, Philadelphia, PA, USA
\\
\bf{2} Department of Physics, Princeton University, Princeton, NJ, USA
\\
\bf{3} Department of Physics, Syracuse University, Syracuse, NY, USA
\\
$\ast$ Present address: Department of Bionanoscience, Delft University of Technology, Delft, The Netherlands; E-mail: t.idema@tudelft.nl\\
$\dag$ E-mail: ajliu@sas.upenn.edu
\end{flushleft}

\section*{Abstract}
Mitosis in the early syncytial \textit{Drosophila} embryo is highly correlated in space and time, as manifested in mitotic wavefronts that propagate across the embryo. In this paper we investigate the idea that the embryo can be considered a mechanically-excitable medium, and that mitotic wavefronts can be understood as nonlinear wavefronts that propagate through this medium. We study the wavefronts via both image analysis of confocal microscopy videos and theoretical models. We find that the mitotic waves travel across the embryo at a well-defined speed that decreases with replication cycle. We find two markers of the wavefront in each cycle, corresponding to the onsets of metaphase and anaphase.  Each of these onsets is followed by displacements of the nuclei that obey the same wavefront pattern. To understand the mitotic wavefronts theoretically we analyze wavefront propagation in excitable media.  We study two classes of models, one with biochemical signaling and one with mechanical signaling. We find that the dependence of wavefront speed on cycle number is most naturally explained by mechanical signaling, and that the entire process suggests a scenario in which biochemical and mechanical signaling are coupled.


\section{Introduction}
The early embryos of many species, including \textit{Drosophila}~\cite{foe83, foe89, foe93},  \textit{Xenopus}~\cite{satoh77,newport82,saka01}, \textit{Oryzias}~\cite{kageyama86}, \textit{Fundulus}~\cite{trinkaus92}, and zebrafish~\cite{kane92,kane93}, exhibit metachronous mitosis, in which mitosis progresses as a wavefront through the embryo.  Such wavefronts are reminiscent of biochemical wavefronts that are used to transmit signals across many cells in other biological systems, such as wavefronts of the molecule cAMP that propagate  in a colony of {\it Dictyostelium} when it begins to aggregate to form a fruiting body~\cite{devreotes89,levine91,lee96}.  Propagating wavefronts, however, need not be purely biochemical in origin.  The process of mitosis is a highly mechanical one that involves significant changes in the volume occupied by chromatin~\cite{kleckner04} as well as separation of chromosomes~\cite{alberts08}. This raises the question of whether mitotic wavefronts are purely biochemical phenomena or whether they might have a mechanical component as well.

The nuclei of the \textit{Drosophila} embryo are syncytial (i.e., they share the same cytoplasm and are not separated into individual cells by plasma membranes) during their first thirteen division cycles. The nuclei  migrate to the egg's surface during the ninth cycle. There they divide five more times, until the fourteenth cycle, when cell membranes form and gastrulation begins~\cite{foe83}.  Mitotic wavefronts are observed in cycles 9 through 13~\cite{foe83}. In this period, chemical diffusion is unhindered by membrane barriers. For example, it is known that calcium, a signal carrier that influences many local phenomena including mitosis~\cite{silver89,silver90,thomas96}, exhibits spikes of concentration in the syncytial embryo~\cite{allbritton93,parry05,parry06,jaffe08,whitaker08} that have been resolved into a wavefront that travels across the embryo at the same speed as the mitotic wavefront~\cite{parry05}.

However, mitosis is also a mechanical phenomenon. In the syncytial embryo, nuclei are embedded in an elastic cytoskeleton, which contains both actin and microtubules~\cite{warn84,karr86,sullivan95}. Actin caps assemble around each of the nuclei at the end of interphase, and provide anchor points for the mitotic spindles that pull the two daughter nuclei apart~\cite{warn84,karr86,sullivan95,stevenson01}. Recent work shows that mechanical interactions are important for re-organization of the nuclei after mitosis~\cite{kanesaki11}, and optical tweezer experiments show that nuclei are mechanically coupled~\cite{schoetz04}. Moreover, mechanical deformations of the embryo are known to be able to induce morphogen expression~\cite{farge03}. However, little is known about how mechanical interactions affect collective phenomena such as mitotic wavefronts at the level of the entire embryo.

In this paper we report the results of both our image analysis of wavefronts in early \textit{Drosophila} embryos, and our theoretical studies of models of wavefront propagation. Using novel tracking techniques, we analyzed confocal microscopy videos taken of \textit{Drosophila} embryos in which the nuclear DNA/chromosomes are visualized by labeling their histones with GFP. Our analysis yields the position, shape and dynamics of the DNA/chromosomes with high temporal and spatial resolution during cycles 9--14. We observe two distinct markers of the mitotic process in each cycle, one corresponding to the onset of metaphase (at which point the chromosomes condense in the nuclear midplane, known as the metaphasic plate, see figure~\ref{fig:cyclestages} for an illustration of the different stages) and one corresponding to the onset of anaphase. Both onsets exhibit identical wavefront patterns, indicating that they are indeed two markers of the same process. Both onsets are also followed by displacements in the positions of the nuclei that also exhibit the same wavefront patterns.  Finally, we find that the wavefront speed slows down from one cycle to the next.

We treat the embryo theoretically as an excitable medium, consisting of nuclei that can be triggered into initiating metaphase or anaphase, thereby locally exciting the medium and thus signaling their neighbors. We not only consider the well-known case of nonlinear wavefront propagation in a chemically excitable medium~\cite{winfree72,agladze82}, but introduce a model for the early embryo as a \emph{mechanically} excitable medium~\cite{idema13prl}, through which mitotic wavefronts can propagate via stress diffusion. Comparing the data with the results of these two models, we find that our observations are difficult to reconcile with a purely biochemical scenario. In such a scenario, the wavefront speed has a tendency to \textit{increase} with nuclear density, and thus with cycle, contrary to our observations. The observations can, however, be explained quite naturally by a novel scenario in which nuclei not only respond to their mechanical environment, but also actively use it to signal each other.  Our results suggest that mitotic wavefronts in syncytial \textit{Drosophila} embryos may constitute one example of a previously unexplored form of mechanical signaling via nonlinear wavefronts that could also arise in very different biological contexts~\cite{idema13prl,majkut13}.

\section{Results}
\subsection{Image analysis results}
\label{sec:observations}
\subsubsection*{Nuclear cycle and shape}
An example image of detected nuclei in a \textit{Drosophila} embryo is shown in figure~\ref{fig:expdata}a. In each cycle, as the nuclei progress from interphase through metaphase to anaphase, the detected shape of the DNA/chromosomes changes in a well-defined manner (figure~\ref{fig:expdata}b). Newly separated nuclei are small and spherical, and thus show up in our shape tracking as small circles. During interphase, the nuclear DNA grows in size over time as it is duplicated. At the onset of metaphase, the chromosomes condense in the midplane of the nucleus, and appear to elongate into an ellipse. The final step of mitosis, the onset of anaphase, corresponds to two detectable changes in the shape: a sudden shift of the orientation axis over a $\pi/2$ angle, and a change of aspect ratio. An example plot showing the ratio of the length of the two axes as a function of time during a cell cycle is given in figure~\ref{fig:expdata}c.

\subsubsection*{Wavefront pattern in the onset of metaphase and anaphase}
The onsets of metaphase and anaphase, as determined by the axes ratio (figure~\ref{fig:expdata}d) are indicated by dotted blue lines and dashed orange lines, respectively. Evidently the onset of metaphase exhibits a wavefront pattern, or rather two wavefronts, one propagating from each pole. The two wavefronts do not necessarily start at the same time. The onset of anaphase exhibits the same wavefront pattern. Mitotic waves were first observed by Foe and Alberts~\cite{foe83}; with better time resolution, it is evident that these wavefronts can be resolved into two distinct markers of mitosis, corresponding to the onsets of metaphase and anaphase. There may well be additional markers that cannot be resolved via histone labeling alone; for example, the work of Parry et al.~\cite{parry05} indicates that calcium may provide another marker for the mitotic process, and we find that the nuclear displacements also provide markers (see below).

\subsubsection*{Effect of shape changes on nuclear positions}
The processes of metaphase and anaphase affect not only the shapes of the chromosomes, but also their positions. After each of the shape changes, the nuclei move \textit{collectively} through the embryo, almost exclusively along the long axis (which we designate as the $x$-axis), resulting in a global `breathing mode' of the entire embryo (see SI movie~1~\cite{movie}). Remarkably, after an initial transition in which the nuclei re-organize after anaphase (studied in detail by Kanesaki et al.~\cite{kanesaki11}), the nuclei hardly move with respect to their nearest neighbors during this collective movement. Figure~\ref{fig:expdata}e shows the average displacement $\Delta x$ along the $x$-axis of a small set of nuclei. Figure~\ref{fig:expdata}f shows the same motion for all nuclei. Note that there are subtle changes in the gray scale that parallel the metaphasic and anaphasic wavefronts but that are shifted to the right (i.e. occur later in time) with respect to each of those wavefronts. This illustrates that the nuclear displacements follow the same wavefront pattern as the axes ratio, so that the displacements also serve as markers for the mitotic wavefront. The existence of such a marker in the displacements as well as in the axes ratio and in calcium concentration underlines the important interplay of mechanics and biochemistry in the mitotic process. 

The displacement response to the onsets of metaphase and anaphase causes the nuclei to move to new equilibrium positions (figure~\ref{fig:expdata}e). Note that the relaxation time of this response is fairly long, about half the length of the mitotic phase ($\sim1\mathrm{min}$) for the onset of metaphase and about half the length of the following interphase (up to $10\mathrm{min}$) for the actual divisions. The displacements following the onset of metaphase therefore occur before the cytoskeletal reconstruction process, which takes place during anaphase, whereas the displacements following the onset of anaphase happen during the aftermath of the cytoskeleton reconstruction.

\subsubsection*{Wavefront speeds}
We quantify the wavefront speeds in figure~\ref{fig:wavefronts} for two sets of movies, where the environmental conditions (in particular the temperature) were approximately the same for all movies in a given set, but differed between the two sets (the data of the two sets were taken several months apart). Figure~\ref{fig:wavefronts}a shows an example of a position vs. time plot of all metaphase (blue diamonds) and anaphase (red pluses) onset events in a single cycle of a single embryo. The slope, corresponding to the wavefront speed, is clearly constant across the embryo. Figure~\ref{fig:wavefronts}b shows the ratio of the speeds of the wavefronts as measured by the onsets of metaphase and anaphase of all embryos, showing that for a given embryo and cycle, these are identical, confirming that they are two markers of a single process. 

From embryo to embryo there are large variations in wavefront speed (figure~\ref{fig:wavefronts}c), but they all show a consistent reduction in speed from one cycle to the next. This trend is illustrated in figure~\ref{fig:wavefronts}e, where we plot the same data, normalized by the speed of the first wavefront, on a log-linear scale. Although our data only span a single decade, this figure suggests that the decrease of wavefront speed with cycle number is consistent with a decaying exponential.

Figure~\ref{fig:wavefronts}d shows that the time interval that separates the onset of metaphase from the onset of anaphase is the same for all cycles for a given embryo, but is different for the two different sets of data. By looking at the point at which the nuclear envelope breaks down and reforms, Foe and Alberts~\cite{foe83} also found that the duration of the mitotic phase is constant through cycles 10, 11 and 12 (3 minutes in their observations, comparable to our result), but was longer for cycle 13 (5 minutes). The re-formation of the nuclear envelope membrane may therefore take significantly longer in the last syncytial cycle, even though the actual mitotic processes continue to follow the pattern of the earlier cycles.

\subsubsection*{Cycle statistics}
The nuclei on the surface are separated by a well-defined distance $a_n$, which decreases with cycle number $n$.  Because the number of nuclei doubles from one cycle to the next, it is not surprising that  $a_n$ decays exponentially, scaling like $a_n \sim 2^{-\beta (n-n_0)}$, with $n$ the cycle number and $n_0$ the number of the first observed cycle. We consistently found a value of $\beta=0.46$ in our experiments (figure~\ref{fig:wavefronts}f and table~\ref{table:expdata}). The value of $\beta$ is slightly less than $1/2$, presumably because the curved embryo is being projected onto a plane. We have also measured the duration of each cycle, $t_n$, and found that, over the observed cycles, it increases with cycle number $n$, with a weak exponential growth: $t_n=t_0 e^{0.29 \cdot n}$, where $t_0=33\mathrm{s}$ for set~1 and $t_0=25\mathrm{s}$ for set~2, see table~\ref{table:expdata} and figure~\ref{fig:cycleduration}.

\begin{figure}[!ht]
\begin{center}
\includegraphics{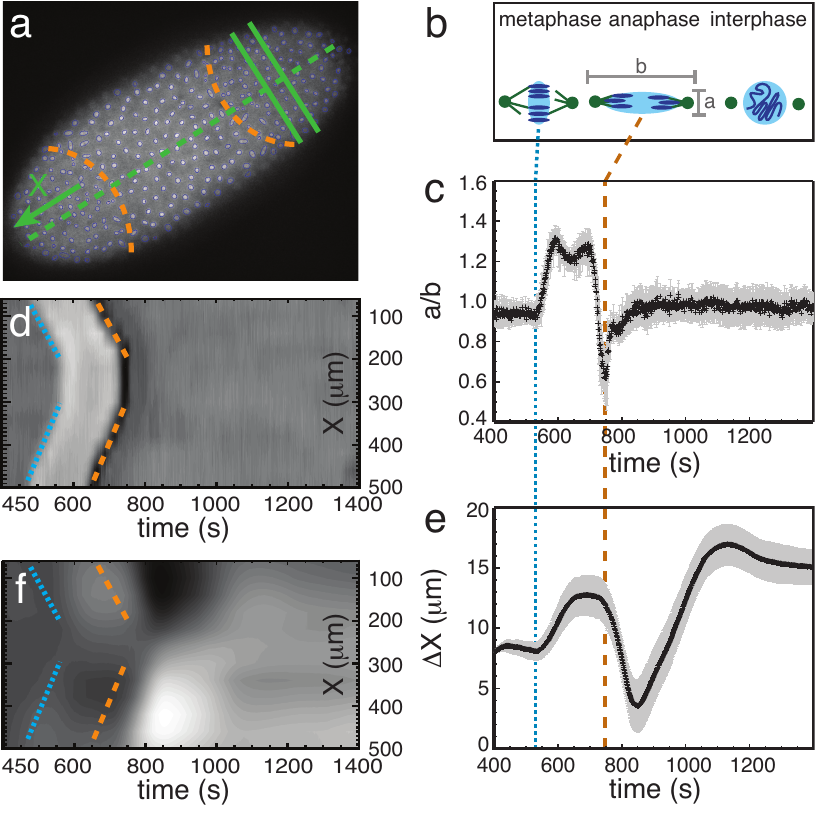}
\end{center}
\caption{{\bf Observation of wavefronts and mechanical response.} a) Image of a \textit{Drosophila} embryo during mitosis at the end of cycle 11, with the detected chromosomal contours overlaid. Anaphasic wavefronts (orange dashed curved lines), the long axis (green dashed straight line) and a typical slice perpendicular to the long axis (green parallel straight lines) are indicated. b) Sketch of the three main states in image analysis: interphase (circular contours), metaphase (compressed elliptical contours), and anaphase (highly extended elliptical contours, perpendicular to metaphase contour). See also figure~\ref{fig:cyclestages}. c) Ratio of the two elliptical axes of the detected shape of the nuclear DNA/chromosomes vs. time in cycle 11, averaged over an $x$-slice (as shown in a); error bars indicate variation within the slice. The transitions between interphase and metaphase, as well as the onset of anaphase, are sharp and indicated respectively by dotted (blue) and dashed (orange) vertical lines. The slice shown was taken at $x=200\mu\mathrm{m}$. d) Kymograph showing the elliptical axes ratio, $a/b$ (where white indicates values larger than 1 and black indicates values smaller than 1), as a function of position $x$ and time. The dotted and dashed lines indicate the onsets of metaphase and anaphase, as in figure~c. e) Average $x$-displacement $\Delta x$ of the nuclei within one slice vs. time. After a nucleus has divided, we use the average position of its two daughters. The slice shown is identical to the one in figure~c. f) Kymograph showing the collective motion of nuclei in slices taken at different positions along the long axis of the embryo. White indicates motion in the positive $x$ direction, black in the negative $x$ direction. Dotted and dashed lines again indicate the onsets of metaphase and anaphase. Note that the displacements occur sometime after these onsets, but follow the same wavefront pattern.}
\label{fig:expdata}
\end{figure}

\begin{figure}[!ht]
\begin{center}
\includegraphics{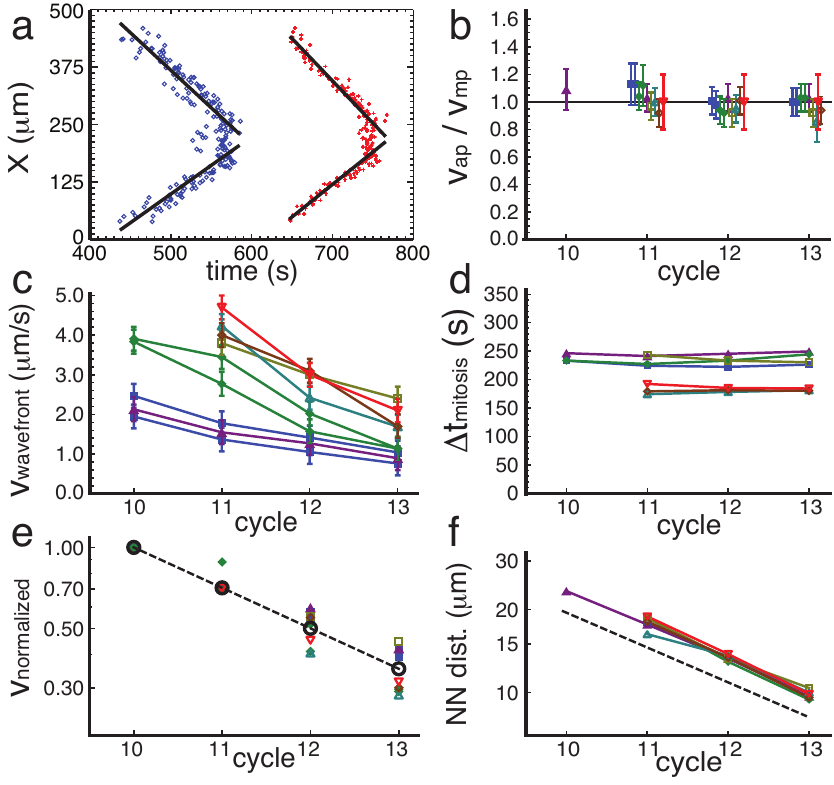}
\end{center}
\caption{{\bf Wavefront propagation and speeds.} a) $x$-coordinate of nuclei at the onset of metaphase (blue diamonds) and anaphase (red pluses) vs. time for the wavefront shown in figure~\ref{fig:expdata}. Both events show two clear wavefronts moving in from near the embryo poles (solid lines). b) Ratio of the speeds of the wavefronts as measured by the onset of anaphase ($v_\mathrm{ap}$) and metaphase ($v_\mathrm{mp}$), for different embryos and cycles. Each embryo is indicated by a different symbol and color, with the closed and open symbols representing two different measurement sets. Ratios for a given cycle and different embryos are slightly separated horizontally. c) Wavefront speed vs. cycle. Two of the embryos contribute two waves per cycle (coming in from opposite poles, as in figure~\ref{fig:expdata}a; blue squares and green diamonds). Although the actual propagation speeds vary significantly from one embryo to the next, they all follow the same trend, decreasing with successive cycles. d) Time interval between the onset of metaphase and anaphase vs. cycle. e) Log-linear plot of wavefront speeds vs. cycle, normalized by the speed of its first observed wavefront (if the first observed wave front is in cycle 10) or 0.71 times its first observed wavefront (if the first observed wavefront is in cycle 11). The black open circles connected by a dashed line corresponds to a scaling of 0.71 per cycle, showing that all embryos follow the same exponentially decaying trend. f) Average distance between nearest neighbors on a logarithmic plot. The dashed line corresponds to a dependence $2^{-\beta n}$, where $n$ is the cycle number and $\beta = 0.46$. In figures b-f, the same symbol/color corresponds to the same embryo.}
\label{fig:wavefronts}
\end{figure}

\subsection{Theoretical Analysis}
Our observation that the mitotic wavefronts propagate at constant speed across the embryo suggests that the embryo can be considered as an excitable medium that supports nonlinear front propagation. Alternatively, the nuclei could all have biological clocks that determine when mitosis starts, which operate independently; in that case the wavefront would be only a result of a lucky timing of those clocks. We discuss various timing models and show that they are inconsistent with our observations in the supplementary material.  Here we concentrate on two distinct classes of models for front propagation in excitable media. In the first model the nuclei communicate by releasing a small chemical species, which then diffuses to neighboring nuclei, triggering them to initiate mitosis. In the second model we explore the novel idea that mitotic wavefronts in the early embryo can be described by wavefront propagation in a medium that is \textit{mechanically} rather than \textit{chemically} excitable. In this model, forces exerted at the onset of the mitotic phase give rise to mechanical stresses that trigger other nuclei to proceed to mitosis as well.

\subsubsection*{Biochemical-signaling model}
At the end of a cycle, when all nuclei have completed the duplication of their DNA, we assume that they are in an excitable state, meaning that they can be triggered to initiate mitosis once they receive an appropriate signal. An obvious candidate for signaling between nuclei is a small protein (\textit{e.g.} a Cdk, cyclin or some other activator), which we will denote as $A$. By definition, nuclei can divide only once per cycle; therefore, in our model, we introduce a refractory period for each nucleus following anaphase, equal to the duration of the interphase.

To introduce chemical excitability, we assume that if the local concentration of $A$ exceeds a threshold $\alpha$, the nucleus starts its program of mitosis, part of which involves releasing more $A$. $A$ then diffuses away, raising the concentration of $A$ at neighboring nuclei, and so on. In our model we allow for a time delay~$t_\mathrm{delay}$ between trigger and release, meaning that a nucleus does not release more $A$ until a time $t_\mathrm{delay}$ after its local concentration exceeds~$\alpha$.  We model releases of $A$ by the nuclei (or sources) as localized pulses (Dirac delta functions), and the system is initiated with a single nucleus releasing a quantity~$Q$ of $A$. The wavefront at any point in time corresponds to the position of all nuclei that release $A$ at that moment. Details on how to solve the diffusion equation and carry out the other needed calculations are given the appendices. An example wavefront is shown in figure~\ref{fig:models}a.

In the case of zero delay time, the speed~$v$ of the resulting wavefront is determined by three parameters: the diffusion constant~$D$, the nuclear spacing~$a$ and the concentration threshold~$\alpha$. We obtained the value of~$a$ from direct measurements (figure~\ref{fig:wavefronts}f). Gregor et al.~\cite{gregor05} found from diffusion experiments in \textit{Drosophila} that the diffusion constant of a molecule with hydrodynamic radius $R$ is well described by a modified Stokes-Einstein relation~\cite{lang86}:
\begin{equation}
\label{eq:modifiedSE}
D=k_{\mbox{\tiny B}} T/(6\pi\eta R) + b,
\end{equation}
where $k_{\mbox{\tiny B}}$ is Boltzmann's constant, $T$ the temperature, $\eta = 4.1\pm 0.4\mathrm{cP}$ the effective viscosity of the syncytial \textit{Drosophila} embryo, and $b = 6.2 \pm 1.0\mu\mathrm{m}^2/\mathrm{s}$ is an experimentally determined constant. Using this expression, we estimate that a reasonable value for the diffusion coefficient (from the size of the activator $A$), would correspond to a chemical with a radius of approximately $5.0\mathrm{nm}$ and therefore a diffusion constant of about $10\mu\mathrm{m}^2/\mathrm{s}$.

Combining the parameters of our model, we define a nondimensional threshold and speed:
\begin{eqnarray}
\label{eq:defbaralpha}
\bar{\alpha} &=& a^2 \alpha / Q, \\
\label{eq:defbarv}
\bar{v} &=& \frac{v}{D/a}.
\end{eqnarray}
In a three-dimensional model the power of $a$ in equation~(\ref{eq:defbaralpha}) is 3. As shown in appendix~\ref{sec:biochemwavefronts}, for a steady-state wavefront, we then have $\bar v = 1/f(\bar \alpha)$, where $f(\bar \alpha)$ increases monotonically with $\bar \alpha$ (figure~\ref{fig:taualphanodelay}). Consequently, if both $D$ and $\alpha$ are fixed, the wavefront speed $v$ increases as the nuclear spacing~$a$ decreases, and thus the speed increases with cell cycle, in direct contradiction to our experimental observations. Thus, the simplest form of the biochemical signaling model cannot describe the data of figure~\ref{fig:wavefronts}c.

We next consider the possibility of a delay~$t_\mathrm{delay}$ between the time when the local concentration of $A$ reaches the threshold value~$\alpha$, and the instant when more $A$ is released. In the limit where $a^2/D \gg t_\mathrm{delay}$, the wavefront speed is determined by diffusion as before, $v = D/ (a f(\bar\alpha))$. In the opposite limit, $a^2/D \ll t_\mathrm{delay}$, we find $v = a / t_\mathrm{delay}$, so $v$ would decrease with cycle number for constant $t_\mathrm{delay}$. We find that for our system, introducing a small, fixed delay time of $2-8\mathrm{s}$ puts us in the crossover regime between these two types of behavior. Consequently, the model predicts that for the first few cycles, the wavefront speed should increase, whereas it should level off or slightly decrease in the last cycle. Changing the value of the threshold~$\alpha$ does not qualitatively change this result. Changing the value of the diffusion constant~$D$ simply shifts the position of the crossover. 

A key result of our analysis with a fixed time delay is that a physically unrealistic diffusion coefficient is required in order to reproduce our experimental observations.  In order to obtain a strictly decreasing wavefront speed for the range of interest, a diffusion constant of more than $100 \mu\mathrm{m}^2/\mathrm{s}$ is required.  This corresponds to a signaling particle that is even smaller than an ion. Thus, a biochemical-signaling model with a time delay that is independent of cell cycle cannot describe our observations either (figure~\ref{fig:models}b).

We also investigated the wavefront speed in the case where the delay time is allowed to vary from one cycle to the next. Naturally, given a value for the diffusion constant and the threshold, for each cycle we can find a delay time such that the speed predicted by the model matches the observed speed; these values are listed in table~\ref{table:diffmodelfits}. The found values do not show any consistent trend, and differ quite strongly between the two data sets. There is no obvious explanation for what would set the time delay in each cycle; the time delay is not proportional to the total duration of the cycle (which increases from one cycle to the next) or any other obvious time scale. Therefore, this procedure simply shifts the problem from understanding the trend in the wavefront speed to understanding the trend in the delay time, and does not provide a satisfactory explanation of our data.

On the basis of these results, we conclude that it is very unlikely that a wavefront that propagates via diffusion of some chemical species would slow down with cycle number, as observed in our experiments.  We also note that any model in which the biochemical signal is mediated by a method that is faster than diffusion (such as active transport) suffers from the same problem: the predicted wavespeed would go up with increasing cycle, because the spacing between the nuclei goes down.

\subsubsection*{Mechanical-signaling model}
The early embryo cannot support ordinary elastic waves because it is heavily damped by the viscosity of the cytosol. Consequently, displacements do not propagate ballistically as in a wave, but diffusively. However, just as diffusion of $A$ can lead to nonlinear wavefront propagation in the biochemical signaling model, diffusion of displacement could lead to wavefront propagation in a mechanical signaling model. We therefore introduce a model in which the nuclei communicate via stresses or strains that they exert on the cytoskeleton at the initiation of the mitotic phase. For example, these could be the forces that cause the chromosomes to condense into sister chromatids in prophase or to align in the nuclear midplane at the onset of metaphase.

In our model, a nucleus starts its  program when the largest eigenvalue of the local stress tensor exceeds a threshold value~$\alpha$. We describe the cytoskeleton as a homogeneous linear elastic medium, characterized by two elastic parameters, for example its bulk and shear moduli ($K$ and $\mu$, respectively) or equivalently the Young's modulus~$E$ and dimensionless Poisson ratio~$\nu$. The viscous fluid in which the elastic cytoskeleton is immersed exerts a drag force on it, characterized by a damping constant~$\Gamma$. Assuming that the nuclei exist in a thin layer near the surface of the embryo, we denote the deformations in the plane of the layer by $u_i = x'_i-x_i$ ($i=1,2$), where the deformation maps point $(x_1, x_2)$ onto point $(x_1', x_2')$. In the overdamped limit (zero Reynolds number), the displacement $\vec u$ of a nucleus can be described by~\cite{chaikinlubensky}:
\begin{equation}
\label{eq:model}
\Gamma \partial_t u_i = \frac{E}{2(1+\nu)} \partial_j \partial_j u_i + \frac{E}{2(1-\nu)} \partial_i \partial_j u_j.
\end{equation}
The term on the left represents the damping with damping factor~$\Gamma$, and the two terms on the right are the elastic force per unit volume. Equation~(\ref{eq:model}) is reminiscent of the diffusion equation: a time derivative on the left equals second-order space derivatives on the right. This model can therefore be thought of as describing the diffusion of the vector displacement field~$u_i$. The right hand side of equation~(\ref{eq:model}) gives rise to two quantities with the dimensions of diffusion constants~\cite{idema13prl}:
\begin{equation}
\label{eq:stressdiffconst}
D_1 = \frac{E}{(1-\nu^2)\Gamma} = \frac{1-\nu}{2} \frac{\mu}{\Gamma} \qquad \mbox{and} \qquad D_2 = \frac{E}{2(1+\nu)\Gamma} = \frac{\mu}{\Gamma}.
\end{equation}

\begin{figure}[!htp]
\begin{center}
\includegraphics[scale=0.75]{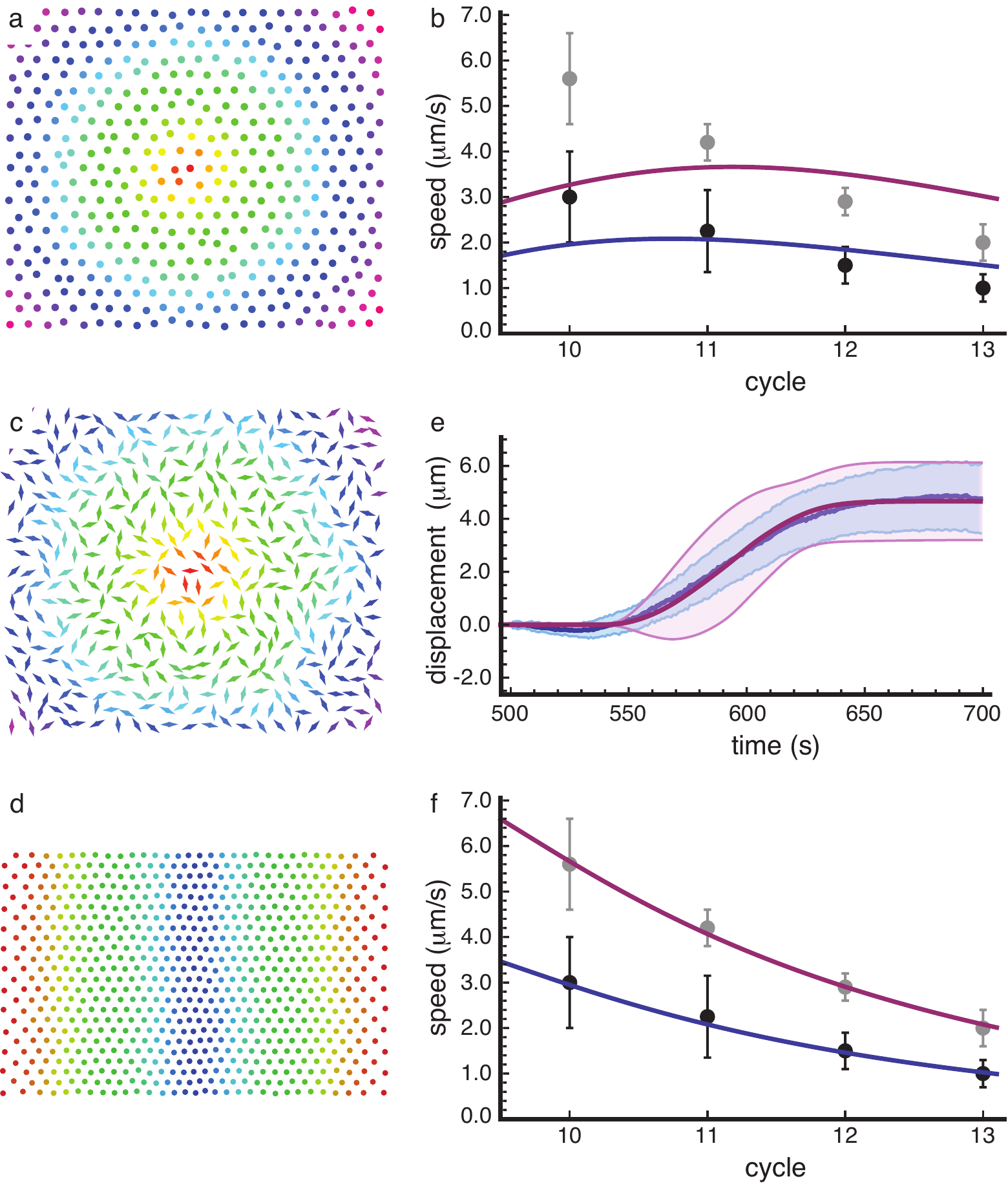}
\end{center}
\caption{{\bf Propagation of wavefronts by chemical and mechanical signaling.} a) Color plot showing the chemical wavefront in two dimensions. The wave starts in the center (red dot) with a single Dirac delta peak release. The color coding indicates when a nucleus releases its chemical to the bulk, going from red through the different hues of the rainbow to violet. b) Plot showing the best fits (blue and purple lines) of the diffusion model with time delay to the to the two sets of experimental data (black and gray dots with error bars). Although the time delay manages to balance the trend that the wavefront speed increases in the region of interest (but not before), the model fails to describe the observed data. Here $D = 10 \mu\mathrm{m}^2/\mathrm{s}$. c) Color plot showing the mechanical wavefront in two dimensions for totally anisotropic dipoles, including their orientations, which are picked at random, and free boundary conditions. The color coding is the same as in figure~a. d) Color plot showing the mechanical wavefront in two dimensions for totally isotropic dipoles and semi-periodic boundary conditions (periodic in vertical direction, free in horizontal direction). Wavefronts are initialized at both free ends simultaneously and travel to the center, as in the experimental system. e) Plot showing fit (purple) of the displacements calculated from the model to the experimentally obtained displacements (blue) following the onset of metaphase. Fit parameters same as in figure~e (set 1). Error bars obtained by averaging over a slice of $40\mu\mathrm{m}$, as indicated in figure~\ref{fig:expdata}a. f) Plot showing fits (blue and purple lines) of the mechanical model for isotropic force dipoles and semi-periodic boundary conditions to the two sets of experimental data (black and gray dots with error bars). Fit parameters: $\alpha = 0.1Q/a_{10}^2$, where $Q$ is the dipole strength and $a_{10}$ the spacing in cycle 10, $\gamma = 1.15$, and $D=3 \mu\mathrm{m}^2/\mathrm{s}$ (blue line/black datapoints), $D=6 \mu\mathrm{m}^2/\mathrm{s}$ (purple line/gray datapoints).}
\label{fig:models}
\end{figure}

In order to introduce \emph{mechanical excitability} into the model, we assume that if the largest eigenvalue of the stress tensor at a nucleus at position $\vec x_0$ exceeds a threshold value, $\alpha$, at time $t_0$, it triggers the nucleus into action which involves adding additional stress to the system. This stress can be added in the form of a source term $Q_{ij} = f_i x_j + x_i f_j$, a symmetric tensor of rank 2, corresponding to a force per unit volume $\vec{f}$ acting over a distance $\vec{x}$. $Q_{ij}$ is therefore the symmetric combination of a force and a distance, with the dimensions of a stress (force per unit area), so it represents a stress source. In two dimensions, $Q_{ij}$ has an isotropic part of the form $Q_0 \delta_{ij}$ and a traceless anisotropic part of the form $Q_1 (n_i n_j - \frac{1}{2} \delta_{ij})$ where $\hat{n}$ indicates the anisotropy direction. If $\hat{n}$ makes an angle~$\theta$ with the $x$-axis, we find that the components of $Q_{ij}$ in matrix notation are given by:
\begin{equation}
\label{eq:sourceterm}
\mathbf{Q} = Q_0 \left(\begin{array}{cc} 1 & 0 \\ 0 & 1 \end{array} \right) \delta(\vec{x})
- Q_1 \left(\begin{array}{cc} \cos 2\theta & \sin 2\theta \\ \sin 2\theta & -\cos 2\theta \end{array} \right) \delta(\vec{x}).
\end{equation}
Here $\delta(\vec x)$ is the two-dimensional Dirac delta function. Similar active force dipoles have previously been introduced into other tissue-level models, such as those of Bischofs et al.~\cite{bischofs04} and Ranft et al.~\cite{ranft10}. To include the force due to the added stress at $\vec{x} = \vec{x}_0$ and $t = t_0$, we add the term $\partial _j Q_{ij}\delta(\vec x - \vec x_0) \Theta (t-t_0)$ to the right hand side of equation~(\ref{eq:model}):
\begin{equation}
\label{eq:modelwithsource}
\Gamma \partial_t u_i =  \frac{E}{2(1+\nu)} \partial_j \partial_j u_i + \frac{E}{2(1-\nu)} \partial_i \partial_j u_j + \partial _j Q_{ij}\delta(\vec x - \vec x_0) \Theta (t-t_0),
\end{equation}
where $\Theta(t)$ is the Heaviside step function. Equation~(\ref{eq:modelwithsource}) essentially describes the diffusion of the vector displacement field~$u_i$ due to a tensor source term. It is similar, but not identical, to a scalar reaction-diffusion equation, which describes the evolution of a scalar concentration field~$c$ due to a scalar source term. It is therefore not surprising that the model described by equation~(\ref{eq:model}) also produces wavefronts, as can be seen in figure~\ref{fig:models}c and d.

In order to compare the model results with the data, we need to estimate the values of the elastic constants and the damping parameter. The speed~$v$ now depends on the quantity $D=\mu/\Gamma$ that determines the dimensional part of both diffusion constants (equation~(\ref{eq:stressdiffconst})), as well as the nuclear spacing~$a$, the strengths~$Q_0$ and $Q_1$ of the source term, and the threshold value~$\alpha$. It is well known that the values of both the elastic and the viscous modulus of a polymer network depend strongly on filament concentration~\cite{doiedwards,muller91,janmey94,palmer99,gardel03,gardel04}, which can differ from one cycle to the next. Because the number of nuclei doubles in each cycle, the number of actin caps in the network doubles as well (see figures~\ref{fig:cyclestages} and \ref{fig:embryocrossection}). Thus, the local concentration of actin and of microtubules should effectively double with cycle number $n$. We therefore write $c \sim 2^{(n-n_0)}$, where as before $n_0$ is the number of the first observed cycle. Both the storage and loss moduli of polymer networks increase with concentration approximately as power laws, but the actual powers are debated~\cite{janmey94,palmer99,gardel03,gardel04}. Moreover, in each successive cycle the nuclei get pushed further out into the plasma membrane encompassing the entire embryo~\cite{foe83}, increasing the friction coefficient. Because the dynamics of our system depend only on the value of the two effective diffusion constants given in equation~(\ref{eq:stressdiffconst}), we will not be able to distinguish the dependence of the storage and loss moduli independently. Instead we assume a dependence $D = \mu/\Gamma \sim c^{-\gamma} \sim 2^{-\gamma(n-n_0)}$.  We will use $\gamma$ as a fit parameter.

Because of the mathematical similarity between the mechanical-signaling model (equation (\ref{eq:model})) and the diffusion model for concentration fields, we can use the same type of dimensional analysis as for the biochemical-signaling model. We again use the dimensionless threshold~$\bar{\alpha}$ and wavefront speed~$\bar{v}$ defined by equations~(\ref{eq:defbaralpha}) and~(\ref{eq:defbarv}), where $Q$ is now the typical strength of the source term, and we write $\bar{v} = g(\bar{\alpha}, \nu)$. We determine $g(\bar{\alpha}, \nu)$ numerically, and find that it can be well described by the functional form $g(\bar{\alpha}, \nu) = -4(c_1+c_2\bar{\alpha})\log(\bar{\alpha}) / (1-\nu^2) + c_3$, where $c_1$, $c_2$ and $c_3$ are constants that depend on the choice of source term and boundary conditions~\cite{idema13prl}. In the analysis that follows, we have adopted boundary conditions that are free along the long axis and periodic along the short axis to mimic the elongated shape of the embryo.

Figure~\ref{fig:models}e shows a fit to a displacement wavefront profile following the first detectable sign of the mitotic wavefront (onset of metaphase) in the initial (tenth) cycle. We find that in order to fit the profile, the source term~(\ref{eq:sourceterm}) must be nearly isotropic, so that $Q_1 \ll Q_0$.  We therefore set $Q_1=0$ and fit to find the threshold stress, which gives $\alpha = 0.1 Q_0/a_{10}^2$, with $a_{10}$ the spacing in cycle 10. Thus, the threshold stress is approximately ten percent of the force exerted per unit area.

Figure~\ref{fig:models}f shows a fit of the wavefront speed of the two datasets, $Q_1=0$ and $\alpha = 0.1 Q_0$.  Here, the fit parameter is the exponent $\gamma$ that governs the change in the displacement diffusion constant from cycle to cycle.  Both datasets are well-described with a value of $\gamma = 1.15$.  The only difference between the two datasets is the value of the displacement diffusion constant $D = \mu/\Gamma$ in the 10th cycle, which is about $3\mu\mathrm{m}^2/s$ for set~1 and about $6\mu\mathrm{m}^2/s$ for set~2. 

These values for the diffusion constant are comparable to those found in microrheology experiments, which have measured the frequency-dependent complex shear modulus in a variety of living cells~\cite{yamada00,fabry01,guigas07,wilhelm08,wirtz09}. In contrast to pure actin networks, living systems often do not exhibit a low-frequency plateau in the storage modulus $G'(\omega)$. Although this makes a precise determination of the shear modulus difficult, we can still get a decent order-of-magnitude estimate from the experimental data at $\mu \sim 5\mathrm{Pa}$. The damping constant $\Gamma$ is given by $\Gamma = c \eta \xi = \eta / \xi^2$~\cite{gardel03,schmidt89}, where $c$ is the filament concentration, $\eta = 4 \cdot 10^{-3} \mathrm{Pa}\cdot{s}$~\cite{gregor05} is the ambient fluid viscosity, and $\xi \sim 100\mathrm{nm}$ is the mesh size of the actin network. We thus estimate $D \sim 10 \mu\mathrm{m}^2/s$, in good agreement with our fitting results. 

The found value for the exponent $\gamma$ is also reasonable. In-vitro experiments on entangled F-actin solutions indicate that the storage and loss moduli depend on the concentration in the same way~\cite{gardel03}, which leads us to expect the shear modulus~$\mu$ and viscosity~$\eta$ to have similar dependence on $c$. On the other hand, for a semidilute solution of rigid rods, the viscosity is expected to rise as $c^3$, where $c$ is the filament concentration~\cite{doiedwards}. Because the damping factor~$\Gamma$  scales with the concentration and the mesh size~$\xi$, which itself depends on the concentration as $\xi \sim c^{-1/3}$, we find that $\gamma$ should be somewhere between $2/3$ (for an entangled F-actin solution) and $8/3$ (for a semidilute solution). Our value of $\gamma = 1.15$ indicates that our system falls somewhere in between these two regimes, which is reasonable for the \textit{Drosophila} embryo, with its hemispherical actin caps enclosing each nucleus (see figure~\ref{fig:embryocrossection}).

Figures~\ref{fig:models}e and f show that we can consistently fit both the wavefront velocity and the displacement profile of the nuclei as a function of time immediately after the metaphasic wavefront, with the same theory.  We note that this is not possible with the chemical signaling model, which cannot provide any information about the displacement profile.  The fact that we can fit both quantities with the same parameters therefore provides strong evidence in favor of the mechanical signaling model.

In addition, we note that the nuclear displacement profile provides a more discriminating test of the mechanical signaling model than the wavefront velocity.   Although the velocity wavefront speed data alone can be fitted by either purely isotropic force dipoles or purely anisotropic force dipoles (and presumably anything in between), the displacement wavefront can only be fit with dipoles with a strong isotropic component. Moreover, although either the displacement or the velocity data can be fit with different combinations of the threshold and diffusion constant, the numbers given above are the only ones for which we can fit both quantities.

In summary, the mechanical signaling model agrees much better with the data than the biochemical signaling model in two important respects.  First, it captures the dependence of the wavefront velocity on cell cycle number while the biochemical signaling model does not.  From dimensional analysis, we have shown for both models that the wavefront velocity depends mainly on $D/a$, where $D$ is the diffusion constant and $a$ is the average distance between nuclei.  Note that $a$ decreases with cycle number.  In the case of biochemical signaling, the chemical diffusion coefficient $D$ remains constant with cycle number, leading to a wavefront velocity that tends to \emph{increase} with cycle number.  In the case of mechanical signaling, however, the displacement diffusion coefficient, $D = \mu/\Gamma$, decreases quite strongly with cycle number because the damping coefficient, $\Gamma$, should increase more rapidly with filament concentration than the elastic constant, $\mu$. If we make the reasonable assumption that the filament concentration increases with cycle number, then this means that the stress diffusion coefficient decreases with cycle number, leading to a wavefront velocity that decreases with cycle number, in accord with experimental observations.  Second, we have shown that the mechanical signaling model describes not only the wavefront velocity but also the displacement profile following the metaphasic wavefront.  In the biochemical-signaling model, a separate mechanical description would be necessary in order to describe the nuclear displacements.

Finally, we note that we have assumed that the elastic constants and damping coefficients vary from cycle to cycle but do not change much during the period that we are focusing on.  However, the cytoskeleton reconstructs completely during the cell cycle.  Our analysis will apply as long as the elastic constants and damping coefficient do not change appreciably from the time that the original triggering wavefront is generated to the time that the anaphasic wavefront occurs.  Thus, the assumption is that cytoskeletal reconstruction occurs sometime during anaphase and is finished before the process of mitosis begins in the next cycle. In particular, this also means that our model should not be able to correctly predict the much larger displacements following anaphase (see figure~\ref{fig:expdata}e), which indeed it cannot.

\section{Discussion}
\label{sec:discussion}
During the early cycles of \textit{Drosophila} development, the cycles of the nuclei are strongly coupled across the entire embryo by mitotic wavefronts that travel at constant speed across the embryo. We summarize our observations as follows:

\begin{enumerate}
	\item There are several markers of the mitotic process in each cycle, corresponding to the onsets of metaphase and anaphase, which are visible as wavefronts that travel across the embryo (figure~\ref{fig:expdata}d).
	\item The speed of the mitotic wavefronts slows down in each successive cycle (figures~\ref{fig:wavefronts}c and~\ref{fig:wavefronts}e).
	\item The onsets of metaphase and anaphase both trigger a mechanical response of the entire embryo in the form of displacements of the nuclei that also exhibit a wavefront pattern (figure~\ref{fig:expdata}f).
	\newcounter{enumii_saved}
      \setcounter{enumii_saved}{\value{enumi}}
\end{enumerate}

In addition to these observations, we add those of Parry et al.~\cite{parry05}:

\begin{enumerate}
      \setcounter{enumi}{\value{enumii_saved}}
	\item There is a visible wavefront in calcium release that coincides with the onset of anaphase.
	\item The speed of the calcium wavefront slows down in each successive cycle, presumably matching the speed of the mitotic wavefront.
\end{enumerate}

We have considered two scenarios to assess whether they are consistent with these observations. In both cases, based on observations (1), (2) and (5), we take the observed metaphase, anaphase and calcium wavefronts to be different markers of the same mitotic process, and assume that the mitotic wavefront is triggered by a single event.

\subsection*{Scenario A} \textit{Mitosis is triggered by a biochemical signal.} Here we assume that a biochemical signal is responsible for triggering mitosis. The signal is mediated by the release and subsequent diffusion of a small ion, molecule or protein. The only chemical species that is known to exhibit a wavefront pattern during mitosis is calcium. However, because the onset of metaphase happens well before the observed calcium wavefront, which coincides with the onset of anaphase (5), calcium cannot be the signal carrier.  Our theoretical analysis suggests that biochemical signaling is unlikely to be consistent with observation (3), since the natural tendency of such a model is to produce a wavefront speed that increases with cycle number.  The larger the signaling molecule, the more pronounced this tendency is. Thus, we conclude that Scenario A is  unlikely.  

This prediction could be tested by looking for wavefronts in likely signaling species.  If the wavefront propagates biochemically, then wavefronts should be observable in the appropriate signaling molecules (presumably CDKs or cyclins that are known to govern checkpoints in the cell cycle that precede the onset of metaphase~\cite{tyson11}).  If, as our model suggests, the wavefront does not propagate biochemically, then the original signaling molecule should not exhibit wavefronts.

\subsection*{Scenario B} \textit{Mitosis is triggered by a mechanical signal.}
In this scenario, there is a mechanical wavefront that triggers mitosis. The signal is transmitted via stress changes in the embryo and amplified by further release of stress as other nuclei enter the mitotic phase. Because this wavefront propagates mechanically, this speed slows down with successive cycles (2).  Since we observe a metaphasic wavefront whose speed of propagation slows down with cycle, the metaphasic wavefront itself could be the triggering mechanical wavefront.  It is more likely, however, that the triggering wavefront occurs earlier in the cycle and starts a clock in each nucleus, which controls the mitotic process.  As a result of this clock, there are many markers of the process that exhibit the same wavefront pattern, including the onsets of metaphase and anaphase (1), the  release of calcium (5), and displacements of the nuclei during metaphase and anaphase (3).  This scenario is consistent with all observations.

Scenario B is consistent as well with independent observations made in \textit{Xenopus} embryos. These embryos are not syncytial; instead they are divided into cells from the first cycle. It is unlikely that a biochemical signal could cross cell membranes to propagate a wavefront. Nevertheless, these embryos do exhibit metachronous mitosis~\cite{satoh77}. They also exhibit calcium oscillations \textit{inside} each cell, which precede anaphase~\cite{keating94}. Their behavior is therefore most consistent with Scenario B: an initial mechanical wavefront triggered by a mechanical process at the onset of metaphase or earlier, is followed by a calcium signal inside each cell and an anaphasic wavefront.  

We emphasize that Scenario B does not imply that the entire process of triggering mitosis is mechanical.  Indeed, the mechanism by which additional stress is generated via a force dipole in our model must be biochemical.  First, there must be some sensor components that are activated when the stress exceeds its threshold value.  These components must then activate other biochemical species to eventually generate additional stress by creating a force dipole.  If Scenario B is correct, there should be a way of incorporating our mechanically signaling model into models of the chemical networks that control the cell cycle, such as those of Tyson and Novak~\cite{tyson11}.  One question is whether the original triggering mechanical wavefront serves as a checkpoint in the cycle.  In order to understand how to include mechanical signaling into such models, it is critical to have new experiments to identify precisely the original triggering wavefront.  Our model would predict that signaling molecules in stages of the cell cycle that follow this triggering wavefront should exhibit wavefronts that slow down with cycle, while those in stages that precede the triggering wavefront should not.

In principle, the estimated elastic constants and damping coefficients could be obtained directly from experiments by measuring the storage and loss moduli of the embryo surface \textit{in vivo} using two-point microrheology. Optical tweezer experiments similar to the ones done by Sch\"otz et al.~\cite{schoetz04} could also be used to extract the elastic moduli and the drag coefficient we used in our mechanical model. The actin concentration could be measured at the same time by staining the actin filaments with e.g. rhodamine, as done by Parry et al.~\cite{parry05} or GFP-moesin, as done by Cao et al.~\cite{cao10}. 

Even though the process of mitosis is known to require chemical activation, the key assumption in Scenario B is that the initial wavefront also \textit{propagates} mechanically.   This can be tested by mechanically poking the embryo at different times within the cell cycle. If the cell is poked just in advance of the original triggering wavefront, the poking itself should generate a wave that propagates from the poking site with the same speed as the mitotic wavefront.  If the embryo is poked too far in advance of the original triggering wavefront, however, there should be no response. If the embryo is poked after the mitotic wavefront begins, there may be no response because the nuclei have already entered mitosis and can no longer be triggered. Thus, we would expect that poking could generate a mitotic wavefront only if it is applied in a certain time window of the cycle that could serve to identify the original triggering wavefront. Note that experiments by Farge at a slightly later stage of development in Drosophila showed that mechanical stress applied in the appropriate time window can lead dramatic changes in development~\cite{farge03}; Scenario B suggests that mechanical stress is important even at the syncytial stages studied here.

Finally, we note that biochemical experiments could also test the mechanical-signaling model.  The most straightforward test would be to to destroy or degrade the filaments that mechanically couple the nuclei.  This should prevent the mechanical wavefronts from propagating and thus the nuclei from synchronizing their mitosis. This could be done by injecting colcemid or nocodazole to disrupt the microtubules or latrunculin which affects actin filaments, for example~\cite{foe93}. Other means of disrupting cytoskeletal filaments, via mutation or laser ablation, should also affect the mechanical wave.

\section{Materials and Methods}
\subsection{Confocal videos}
The imaged flies were from a His-GFP stock with a P[w+ ubi-H2A-GFP] insertion on the third chromosome. All embryos were collected at $25^\circ\mathrm{C}$ and dechorionated in 100\% bleach for 1 minute. They were picked using a $70\mu\mathrm{m}$ nylon strainer (BD Falcon), rinsed in distilled water and laid down on a semipermeable membrane (Biofolie). The excess water was absorbed and the embryos were immersed in Halocarbon oil 27 (Sigma Aldrich) and covered with a $22\times22\mu\mathrm{m}$ coverslip (Corning). Embryos were imaged with a $20\times$ oil immersion objective plan apochromat (Leica, NA=0.7) on a Leica SP5 laser scanning microscope with excitation wavelength of $488\mathrm{nm}$ (argon laser $60\mathrm{mW}$). 8~bit images were taken every second at $512\times1024$ $0.45\mathrm{nm}$ pixels and $1.4\mu\mathrm{s}/\mathrm{pixel}$ ($734\mathrm{ms}/\mathrm{image}$). An example video is shown in SI movie~1~\cite{movie}.

\subsection{Image analysis}
We visualized nuclear DNA/chromosomes by tagging their histones with GFP. To determine the positions, sizes, aspect ratios and orientations of the DNA/chromosomes from each video frame, we developed a new image analysis technique, explained in detail in~\cite{idema13ebj}. In brief, we first applied a bandpass filter to eliminate high-frequency noise. We then made a contour plot of the resulting image, found the locally highest-level contour (i.e., the contours with no other contour inside them), and identified each of them as a single nucleus. For each nucleus, we fit the contour at half-height with an ellipse to get its  position, shape and orientation. An example of an experimental image with the chromosomal tracking overlaid is given in figure~\ref{fig:expdata}a.

Because the images were taken at high frequency (typically 1~Hz), the nuclei move less than their own radius from one frame to the next, simplifying tracking. The obvious exception is when nuclei divide during anaphase, and the observed shape splits in two. Because we detect shapes as well as positions of the chromosomes in each nucleus, tracking divisions is easy as well: when a nucleus divides, the chromosomes become highly elongated just before they split, and produce two almost circular daughters close to the endpoints of the long axis of the mother immediately after it splits, which are easily identified.

\section*{Acknowledgments}
We thank Thomas Gregor for providing resources for the experiments and for his careful reading of the manuscript, and Xiaoyang Long for assistance with acquiring the experimental data. We also thank Gareth Alexander and Michael Lampson for helpful discussions. This work was partially supported by the Netherlands Organization for Scientific Research through a Rubicon grant (T.I.) and by the National Science Foundation through grants NSF-DMR-1104637 (A.J.L.) and NSF-DMR-1104707 (L.K. and T.C.L.).

\section*{Appendices}
\appendix
\section{Embryo layout and replication cycle}
\begin{figure}[hptb]
\begin{center}
\includegraphics[width=0.5\columnwidth]{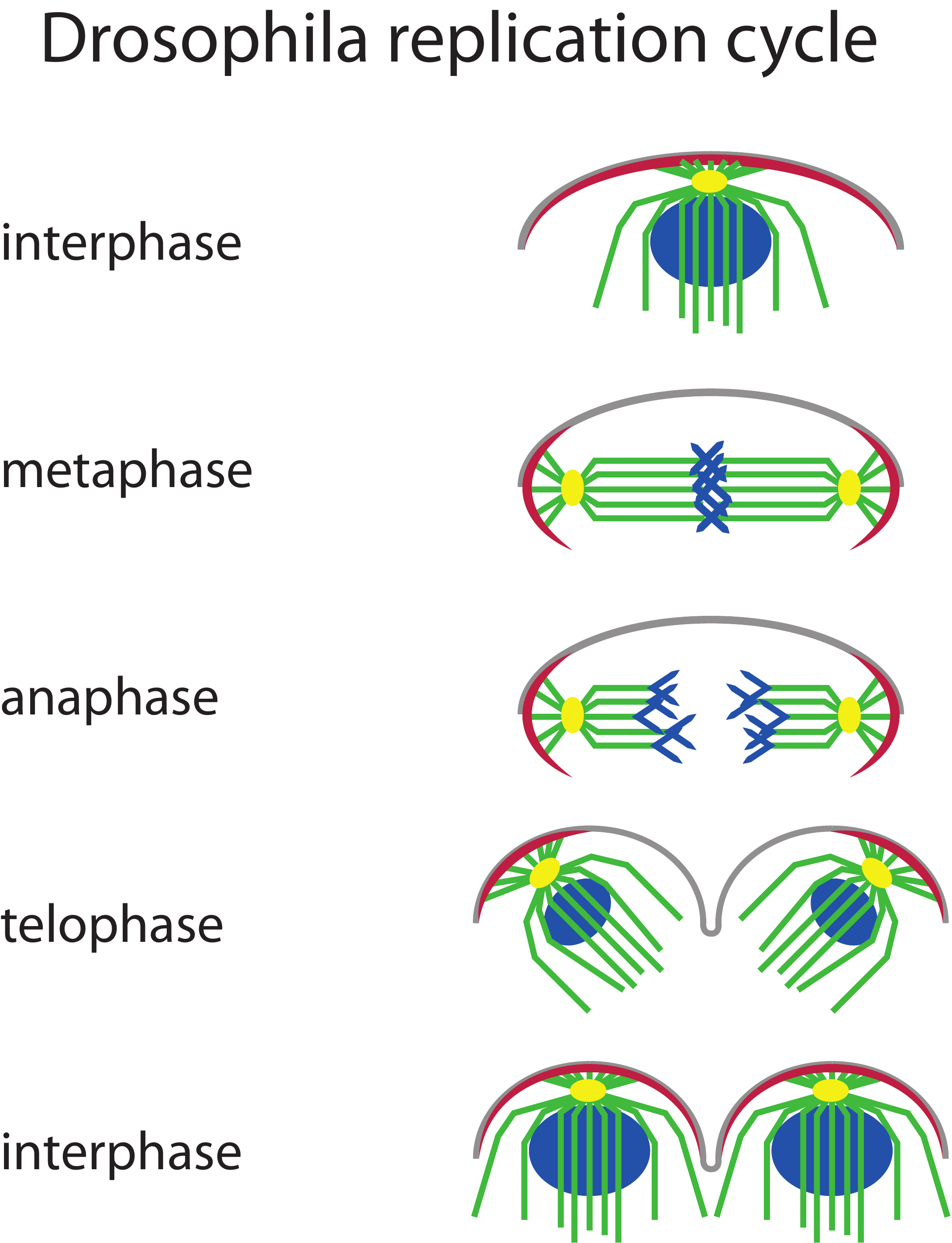}
\end{center}
\caption{Illustration showing the four stages of the \textit{Drosophila} embryo replication cycle that we can detect from our movies: interphase (DNA replication), metaphase (condensation of chromosomes in the nuclear midplane), anaphase (division of the nucleus in two daughter nuclei) and telophase (separation of daughter nuclei). The plasma membrane is shown in gray, the actin cap (made of actin filaments) in red, the microtubules in green, the centrosomes in yellow, and the DNA/chromosomes in blue.}
\label{fig:cyclestages}
\end{figure}

\begin{figure}[hptb]
\begin{center}
\includegraphics[width=\columnwidth]{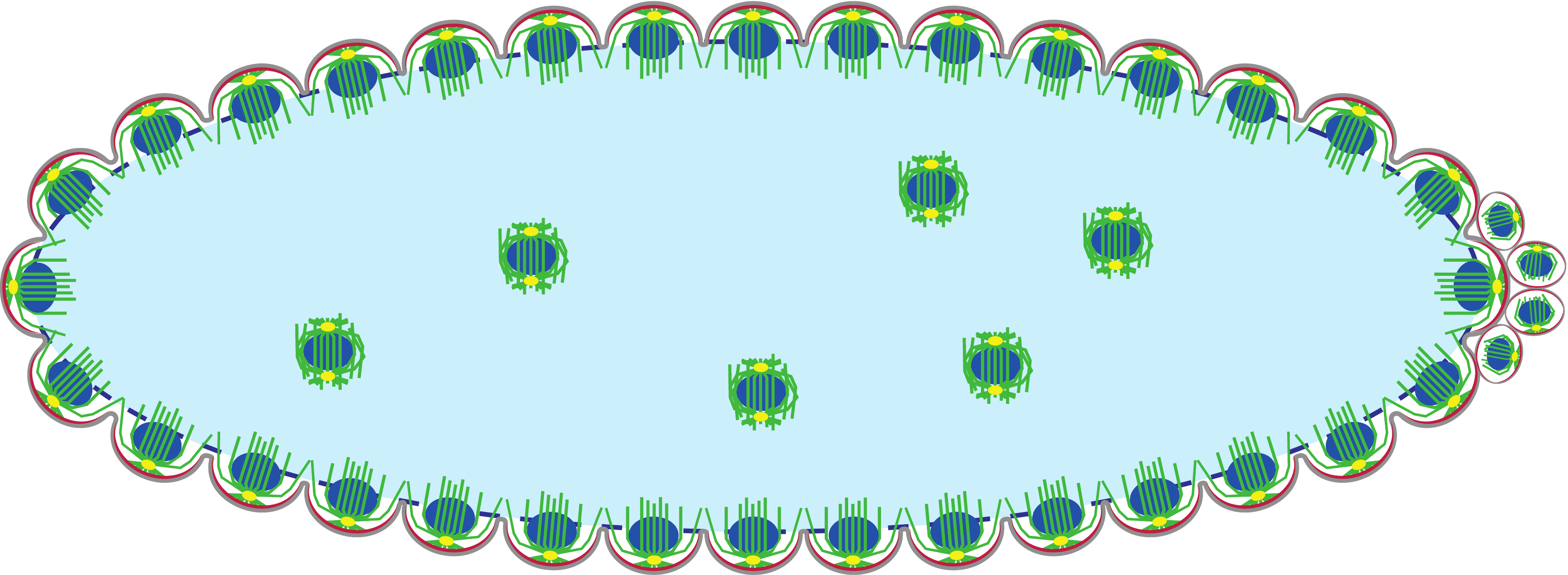}
\end{center}
\caption{Sketch of a cross-section through a \textit{Drosophila} embryo valid for stages 9-13. Most nuclei are located at the surface of the embryo. The nuclei are pushed outwards into the plasma membrane (gray), resulting in the formation of somatic buds. Each nucleus is enclosed in a microtubule basket (green) and contained in an individual actin cap (red), which gets disassembled after mitosis and re-assembled during interphase. DNA/chromosomes are shown in blue and centrosomes in yellow. The yolk (light blue) is a viscoelastic fluid containing water, cytoskeletal elements and necessary building blocks for the nuclei. The yolk is bounded by an actin cortex over which the nuclei can move. Also shown in this sketch are the small number of nuclei that reside inside the yolk, and the also small number of somatic cells that already form in cycle 10 at the posterior end (the pole cells that divide out of sync with the rest of the embryo). See Foe and Alberts~\cite{foe83} for sketches for each of the first 14 cycles and Schejter and Wieschaus~\cite{schejter93} for a review on the cytoskeletal elements in the early embryo.}
\label{fig:embryocrossection}
\end{figure}

The four main stages in the \textit{Drosophila} embryo replication cycle, which we can detect from our movies, are illustrated in figure~\ref{fig:cyclestages}. A sketch of a cross-section of the embryo is shown in figure~\ref{fig:embryocrossection}, illustrating how the nuclei are all located at the surface of the embryo for cycles 9-13~\cite{foe83,schejter93}.

\section{Experimental data sets}
Our image analysis results are for two different sets of experiments, which were carried out at ambient room temperature several months apart.  The ambient temperature was higher for the second set, resulting in faster embryo development.  We only used the data from those embryos which we could track from cycle 10-14 in the first set (Dataset 1, 3 embryos) and cycle 11-14 in the second set (Dataset 2, 4 embryos). SI movie~1~\cite{movie} is the raw data of one of the embryos from set~1. This confocal microscopy imaging movie shows a developing Drosophila embryo. The chromosomal histones are visualized by labeling with GFP. The version of the movie shown here shows 1 image per 15s, displayed at 5fps, so sped up 75x. Movies for data analysis were recorded at 1fps. The dimensions of each frame are $346\times 440\mu\mathrm{m}$.

\section{Additional image analysis results}
\begin{figure}[hptb]
\begin{center}
\includegraphics{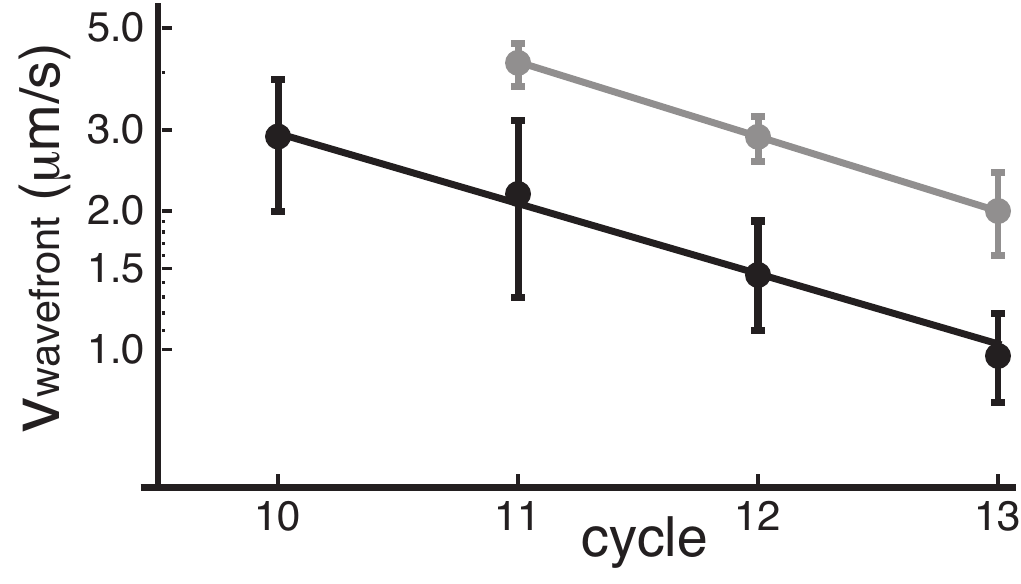}
\end{center}
\caption{Average speed of each of the two sets of data, on a log-linear plot, fitted by an exponential $v \sim 2^{-\varepsilon (n-n_0)}$, $\varepsilon = 0.5 \pm 0.05$. The black dots correspond to the mean wavefront speeds of set~1, and the gray ones to the mean speeds of set~2.}
\label{fig:speeddataexpfit}
\end{figure}

The average data from the two sets are given in table~\ref{table:expdata}, and their average speeds are plotted on a log-linear scale in figure~\ref{fig:speeddataexpfit}. The data from set~1 are given as closed symbols (blue, purple and green) in figure~2 of the main text, the data from set~2 as open symbols (cyan, orange, gold and red). In figure~3c of the main text and figure~\ref{fig:speeddataexpfit}, the black dots correspond to the mean wavefront speeds of set~1, and the gray ones to the mean speeds of set~2.

\begin{figure}[hptb]
\begin{center}
\includegraphics{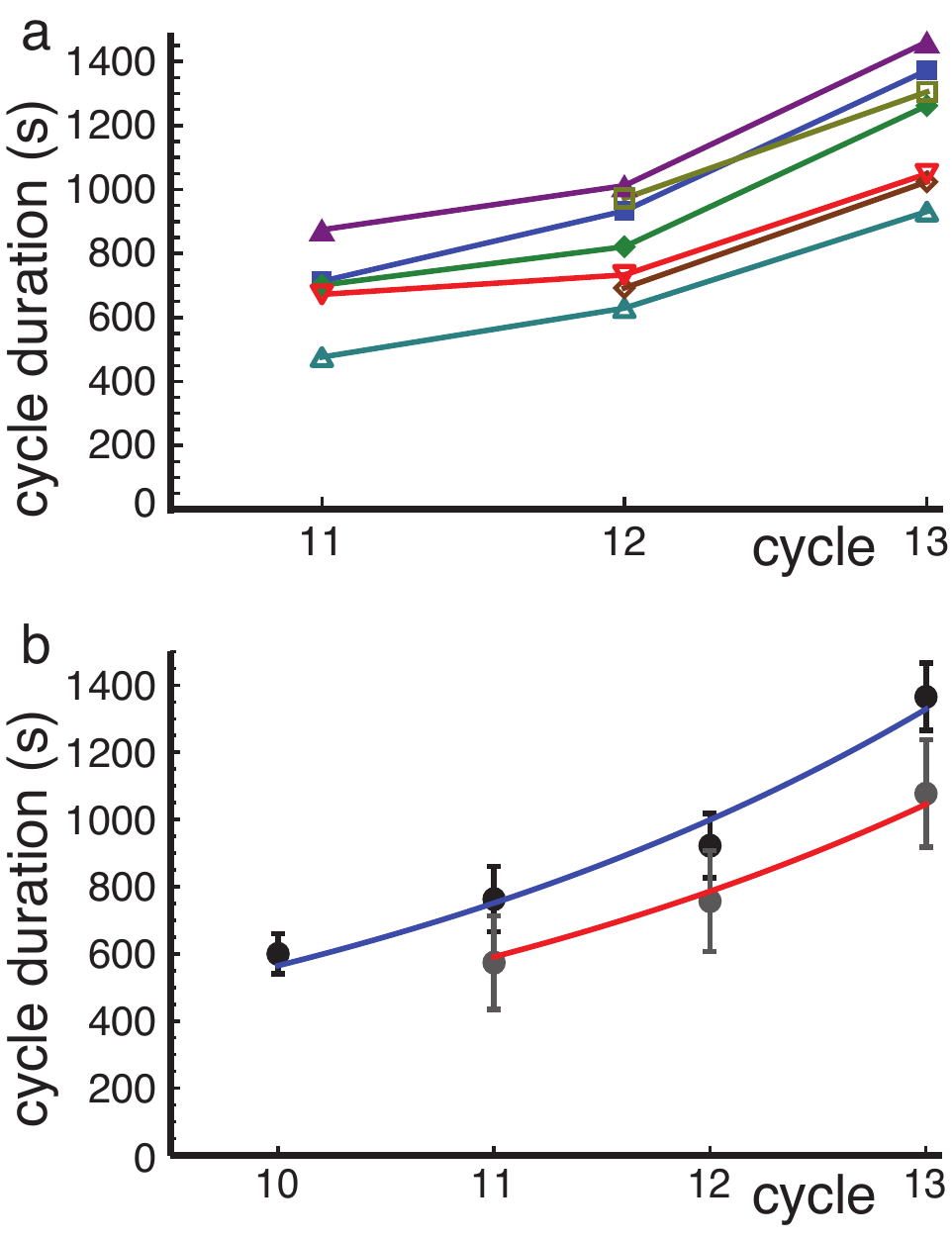}
\end{center}
\caption{Duration of the measured cycles. a) Experimental data. The different symbols and colors correspond to the ones in figure~2 in the main manuscript. b) Cycle duration averaged over all experimentally observed embryos (black and gray dots for sets~1 and 2 respectively). The cycle durations can be fitted reasonably well by a weak exponential $t_n = t_0 e^{0.29 \cdot n}$, where $t_0 = 33\mathrm{s}$ (set~1) and $t_0 = 25\mathrm{s}$ (set~2).}
\label{fig:cycleduration}
\end{figure}

In addition to the data shown in figure~2 of the main text, we also measured the duration of each of the cycles (figure~\ref{fig:cycleduration}a). The numbers we found are consistent with those reported by Foe and Alberts~\cite{foe83} and Parry et al.~\cite{parry05}. Averaging over the embryos in each set, we find that the cycle duration can be reasonably approximated by a quadratic dependence on the cycle number (figure~\ref{fig:cycleduration}b).

\begin{table}[hp]
\begin{center}
\begin{tabular}{|c|c|c|c|c|}
\hline
\multicolumn{5}{|c|}{Data set 1} \\
\hline
cycle number & 10 & 11 & 12 & 13 \\
\hline
nuclear spacing ($\mu\mathrm{m}$) & $23.4\pm0.8$ & $18.2\pm0.6$ & $13.2\pm0.3$ & $9.7\pm0.2$ \\
wavefront speed ($\mu\mathrm{m}/\mathrm{s}$) & $2.9\pm0.9$ & $2.2\pm0.9$ & $1.5\pm0.4$ & $1.0\pm0.2$ \\
cycle duration (s) & $600\pm60$ & $763\pm97$ & $922\pm96$ & $1365\pm100$ \\
mitosis duration (s) & $237\pm8$ & $231\pm9$ & $233\pm12$ & $240\pm12$ \\
\hline
\end{tabular}

\begin{tabular}{|c|c|c|c|}
\hline
\multicolumn{4}{|c|}{Data set 2} \\
\hline
cycle number & 11 & 12 & 13 \\
\hline
nuclear spacing ($\mu\mathrm{m}$) & $18.0\pm1.2$ & $13.5\pm0.3$ & $10.0\pm0.4$ \\
wavefront speed ($\mu\mathrm{m}/\mathrm{s}$) & $4.2\pm0.4$ & $2.9\pm0.3$ & $2.0\pm0.4$ \\
cycle duration (s) & $574\pm139$ & $757\pm150$ & $1077\pm160$ \\
mitosis duration (s) & $197\pm32$ & $194\pm26$ & $194\pm24$ \\
\hline
\end{tabular}
\end{center}
\caption{Experimental data averaged over the data sets. Data sets 1 and 2 correspond to two different sets of measurements, taken on different days. They correspond to respectively the closed and open symbols in figure~2 of the main text and figure~\ref{fig:cycleduration}a.}
\label{table:expdata}
\end{table}

\newpage

\section{Timing models}
Consider the following timing mechanism for generating a wavefront in a row of people. Assume each person has a (synchronized) watch, and each is told to raise his/her arms at exactly $\tau$ seconds, according to their individual clocks, after an arbitrarily chosen $t=0$. If all of the clocks run at the same rate, there is no wave, but if the clock of each successive person in the line runs more slowly that that of his/her neighbor to the right, then a wavefront will be generated.

For the timing method to be the cause for the wavefronts observed in our system, each nucleus would require a clock. That clock could simply be the amount of time it takes to duplicate the DNA, i.e., the duration of the interphase, which changes from one cycle to the next. However, there is no correlation between a nucleus' position and the duration of interphase, which means that we would not expect a mitotic wavefront to emerge in this case. Alternatively, it is well-established that there are several proteins which exhibit patterning along the anterior-posterior or dorsal-ventral axes of the embryo. A much studied example is Bicoid, which exhibits an exponential profile along the anterior-posterior axis~\cite{driever88a,gregor07}, the same axis along which the wavefront travels. Now if the duration of interphase were affected by the local Bicoid concentration, that could provide a mechanism for the clocks of the nuclei to get out of sync, and produce a wavefront in the various markers for mitosis (such as our observed metaphase and anaphase wavefronts).

There are three reasons why the model outlined above cannot explain our data. The first is specific to Bicoid. As observed by Gregor et al.~\cite{gregor07}, the total amount of Bicoid steadily increases over time, as more of the protein is translated in each cycle. In particular, the amount of protein keeps steady pace with the number of nuclei, such that at the start of each cycle, the actual amount of protein in each nucleus at a given position in the embryo is always the same. Therefore if Bicoid were responsible for causing the mitotic wavefronts, the wavefronts would have the same speed in each cycle, which they do not. The second reason is more general: as we show below, in order to obtain a linear wavefront propagation from an exponential concentration profile, the actual absolute amount of material does not matter, only the decay length - which means that once again the predicted wavefront speed would be independent of cycle. Finally note that to get wavefronts traveling in both directions along the anterior-posterior axis of the embryo, we would need at least two concentration gradients of different proteins, for which it would be highly surprising if they produced wavefronts with the same speed. The timing method therefore cannot describe our data.

We now assume that the nuclei do \emph{not} signal each other, in any way, that it is time to start mitosis; instead, they sample their local environment for a given protein, such as Bicoid (Bcd), and have the length of their cycle depend on the local concentration. Because the concentration of many such proteins does indeed exhibit a gradient from one of the two poles, this could explain how mitosis starts close to the poles, and then seems to `travel' along the embryo, where in reality there is no traveling front at all.

Roughly speaking, there are four kinds of patterns expressed by proteins in the early \textit{Drosophila} embryo: an exponential profile along the AP axis starting from one of the poles (of which Bcd is an example~\cite{driever88a,gregor07}), an exponential profile along the DV axis (such as Dorsal~\cite{sample10}), a terminal morphogen profile which is high at both poles and both low and flat in between (as for the phosphorelation gradient of MAPK~\cite{sample10}), and a striped pattern (for e.g. Hunchback, Giant, Paired and Runt~\cite{gregor05}). Because the observed mitotic waves start at the poles and then spread along the AP axis of the embryo, the striped patterns and the DV-axis gradients cannot be the ones causing them. There are often \emph{two} mitotic waves, which start at the two opposite poles, which suggest that the terminal morphogens might be good candidates, but they hardly show any profile in the mid-60\% of the embryo~\cite{sample10}. The most likely candidates are therefore the proteins that have an exponential profile along the AP axis, although in this case there must be at least two proteins that can trigger mitosis. This latter observation is a first weak point of the timing model, but not necessarily cause to rule it out.

During interphase, the nuclei in the syncytial embryo are surrounded by a nuclear membrane (also known as the nuclear envelope). One of the things the nucleus can do is to concentrate proteins inside that membrane. This has been observed directly for Bcd by Gregor et al.~\cite{gregor07} and confirmed by our own observations. Irrespective of whether the proteins are concentrated in the nuclei during interphase, or they are distributed throughout the entire cytosol, they always exhibit an exponential decay along the anterior-posterior axis. As observed by Gregor et al.~\cite{gregor07}, the total amount of Bcd steadily increases over time, as more of the protein is translated in each cycle. In particular, the amount of protein keeps steady pace with the number of nuclei, such that at the start of each cycle, the actual amount of protein in each nucleus at a given position in the embryo is always the same~\cite{gregor07}.

Let us denote the position along the long axis by~$x$, and the total length of the axis by~$L$. The local concentration at $x$ is then given by $c(x) = c_0 e^{-x/\lambda}$, where~$\lambda$ is the characteristic length scale of the exponential profile. As stated above, the experimental results of Gregor et al. tell us that $\lambda$ is the same in all cycles, whereas $c_0$ goes up~\cite{gregor07}. Assuming that all nuclei are equally good at collecting material from their environment, the amount of material collected by a single nucleus in a simple one-dimensional model of the embryo is given by
\begin{equation}
\label{totalstuff}
C(x, N) = \int_{x-L/2N}^{x+L/2N} c(y) \dd y = 2 c_0 \lambda \sinh\left(\frac{L}{2 N \lambda}\right) e^{-x/\lambda},
\end{equation}
where $N$ is the number of nuclei. As stated above, the key assumption of the timing model is that the duration of the cycle of each nucleus depends somehow on the concentration, or rather, on the amount of material in the nucleus, so we have
\begin{equation}
\label{cycleduration}
\Delta t_\mathrm{cycle} = f( C(x, N)) = f\left( 2 c_0 \lambda \sinh\left(\frac{L}{2 N \lambda}\right) e^{-x/\lambda} \right).
\end{equation}

Unfortunately, we do not know what the function~$f$ in equation~(\ref{cycleduration}) is. The only thing we do know is that it is monotonously increasing with its argument (the total amount of material in a nucleus). We will therefore explore two explicit possibilities:
\begin{itemize}
\item The simplest possible dependence: $f$ is a linear function.
\item The dependence that gives the observed behavior of (effective) wavefronts, i.e., that the resulting `speed'~$v$ of mitosis events through the embryo is well-defined and constant throughout, and $\Delta t_\mathrm{cycle} = x/v$.
\end{itemize}

For the first option, we write $f(\alpha) = t_0 - \tau \alpha$, with $t_0$ some offset time and $\tau$ a timescale. We can then calculate the speed of an observed mitotic wavefront, as a function of the number of nuclei~$N$, by calculating the time difference between two positions $x$ and $y$ in the embryo:
\begin{eqnarray}
\label{vlinearmodel}
v(x, y, N) &=& \frac{y-x}{\Delta t_\mathrm{cycle} (y, N) - \Delta t_\mathrm{cycle} (x, N)} \nonumber \\
&=& \frac{y-x}{2 c_0 \lambda \tau \sinh\left(\frac{L}{2 N \lambda}\right) \left( e^{-x/\lambda} - e^{-y/\lambda}\right)}.
\end{eqnarray}
Equation~(\ref{vlinearmodel}) shows that $v$ depends on the position, so there is no well-defined wavespeed in this model. This is of course no big surprise - we just took a random functional dependence for $f$, so there should be no reason to expect it would produce a wavespeed that is position-independent. However, this does illustrate the point that a constant wavespeed is something special: we need to specifically \emph{choose} $f$ such that a constant speed comes out.

Note that it may of course be that there is no constant wavespeed, but that it only appears to be constant within our error bars. Although we can not rule this option out, this also does not come out naturally. For instance, inserting numbers for \textit{Drosophila} from Gregor et al.~\cite{gregor05} ($L = 450 \mu\mathrm{m}$, $\lambda = 70 \mu\mathrm{m} = 0.15 L$, $N=50$) we find that for $x=0$, the measured speed more than doubles as we take the measuring point $y$ across the embryo, which can certainly not be confused with a constant speed.

To get a constant speed, we need to choose a different function~$f$, specifically a logarithm: $f(\alpha) = t_0 + \tau \log \alpha$. In this case we find:
\begin{equation}
\label{vlogmodel}
v(x, y, N) = \frac{y-x}{(\tau/\lambda) (y-x)} = \lambda/\tau.
\end{equation}
In this case, we do indeed find a constant value of $v$ across the embryo. However, we also find that $v$ is independent of $N$. $v$ does depend on the decay length~$\lambda$, but the value of~$\lambda$ does not change~\cite{gregor07}. The wavefront speed predicted by this model is therefore the same for all cycles, in direct contradiction to our observations.

\section{Diffusion model}
The process of diffusion is governed by the diffusion equation, here given for a concentration field~$c(\vec x,t)$:
\begin{equation}
\label{diffeq}
\diff{c(\vec x,t)}{t} = D \nabla^2 c(\vec x,t),
\end{equation}
where $D$ is the diffusion constant. Equation~(\ref{diffeq}) is linear, so we can use the superposition principle: the sum of any two solutions is itself a solution. The general solution for a system with no boundaries depends only on the initial condition~$c(\vec x, 0)$, and is given by
\begin{equation}
\label{DEsol}
c(\vec x,t) = \int G(\vec x, \vec y, t) c(\vec y, 0) \dd \vec y,
\end{equation}
where $G(\vec x, \vec y, t)$ is the Green's function of the diffusion equation, which for a system in $n$ dimensions is given by
\begin{equation}
\label{DEGF}
G(\vec x, \vec y, t) = \frac{1}{(4 \pi D t)^{n/2}} \exp\left(-\frac{|\vec x - \vec y|^2}{4 D t}\right).
\end{equation}
The Green's function describes the concentration field at $\vec x$ at time $t$ due to a single delta-function concentration source at $\vec y$ at time $0$.

\section{Mechanical model}
\begin{figure}[htb]
\begin{center}
\includegraphics{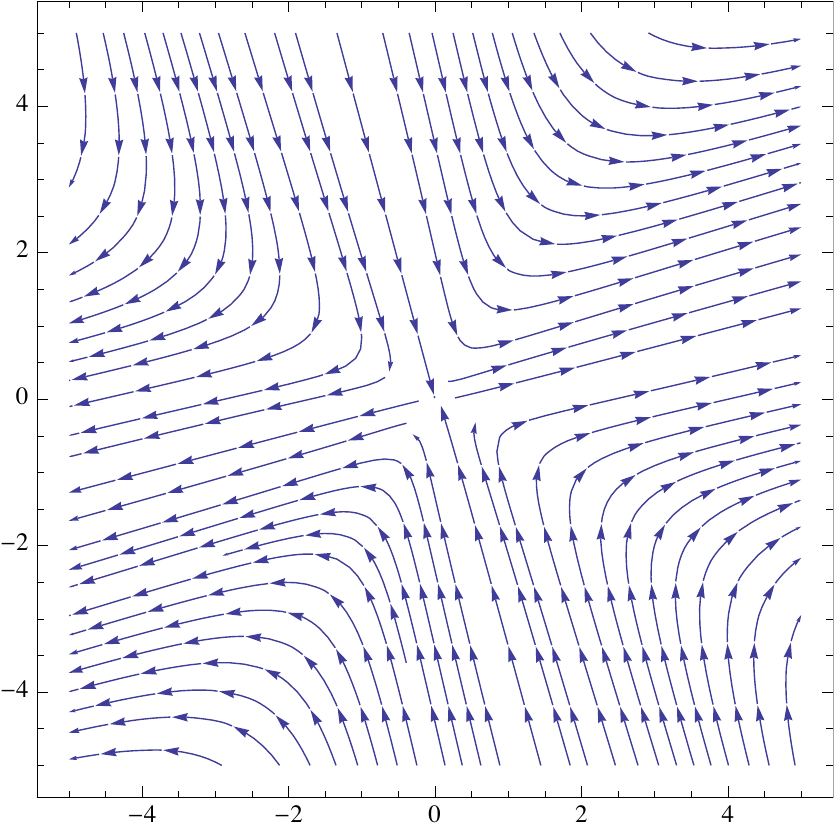}
\end{center}
\caption{Example of a displacement field due to a single force dipole located at the origin and having an angle of $\pi/6$ with respect to the $x$-axis, obtained by taking the $t\to\infty$ limit of $G_{ijk} (\vec x, t) Q_{ij}$.}
\label{fig:statdipole}
\end{figure}

As described in the main text, we can describe the medium in which the nuclei live as an overdamped elastic medium. Motion in this medium can be described by some displacement vector~$\vec u$ from a fixed reference position. We get force balance by equating the damping forces acting on the nuclei (due to friction with the cortical actin layer surrounding the yolk or the outer membrane, and drag due to the viscous fluids the nuclei and their surrounding microtubule baskets are immersed in) to the elastic forces in the polymer cytoskeleton:
\begin{equation}
\label{mechanicalmodel}
\Gamma \partial_t u_i = \frac{E}{2(1+\nu)} \partial_j \partial_j u_i + \frac{E}{2(1-\nu)} \partial_i \partial_j u_j.
\end{equation}
As also pointed out in the main text, equation~(\ref{mechanicalmodel}) is reminiscent of the diffusion equation: a time derivative on the left equals second-order space derivatives plus a source term on the right, and it comes as no surprise that the solution depends on a quantity with the dimension of a diffusion constant $D = \mu / \Gamma$, where $\mu$ is the material's shear modulus and equals $E/(2+2\nu)$. Moreover, equation~(\ref{mechanicalmodel}) also allows for a Green's function type solution, but here in the form of a tensor $G_{ijk}(\vec x, t)$, relating an arbitrary input~$Q_{ij} \delta(\vec x) \Theta(t)$ to a resulting displacement vector $u_k(\vec x, t)$~\cite{idema13prl}. In two dimensions, the input tensor $Q_{ij}$ has three independent components, and can be decomposed in a (hydrostatic) expansion/contraction, and two symmetric traceless parts:
\begin{equation}
\label{eq:sourceterm1}
\mathbf{Q} = Q_0 \left(\begin{array}{cc} 1 & 0 \\ 0 & 1 \end{array} \right)
+ Q_1 \left(\begin{array}{cc} 1 & 0 \\ 0 & -1 \end{array} \right)
+ Q_2 \left(\begin{array}{cc} 0 & 1 \\ 1 & 0 \end{array} \right).
\end{equation}
The two symmetric traceless parts can be converted into each other by a rotation over $\pi/4$. They can therefore alternatively be written as a single term with a prefactor and an angle, as done in equation~(3) of the main text:
\begin{equation}
\label{eq:sourceterm2}
\mathbf{Q} = Q_0 \left(\begin{array}{cc} 1 & 0 \\ 0 & 1 \end{array} \right)
- Q_1 \left(\begin{array}{cc} \cos 2\theta & \sin 2\theta \\ \sin 2\theta & -\cos 2\theta \end{array} \right).
\end{equation}
The second term in equation~(\ref{eq:sourceterm2}) now represents a volume-conserving force dipole that is oriented at an angle~$\theta$ with respect to the $x$-axis. An example displacement field due to a single force dipole at the origin is given in figure~\ref{fig:statdipole}.

\section{Wavefronts}
Now that we know the solutions to the chemical and mechanical diffusion equations due to a single source, we can exploit the fact that the chemical and stress diffusion equations~(\ref{diffeq} and \ref{mechanicalmodel}) are linear to compute the behavior of a system with many sources using the superposition principle. For simplicity, we pre-arrange the nuclei on a triangular grid, with a little noise in the position of each nucleus to prevent artifacts due to a perfect arrangement. This is consistent with our observation that just before the mitosis waves the nuclei in an actual embryo have a rather high degree of triangular order, except where there are defects due to the fact that a nucleus did not divide in an earlier cycle. Alternatively, we can also consider packings with short-range correlations but no long-range order (like the packing of soft repulsive spheres), which gives the same results~\cite{idema13prl}.

\subsection{Wavefronts in the biochemical diffusion model}
\label{sec:biochemwavefronts}
We will describe the release by a nucleus of a biochemical with a Dirac delta function source located at the position of the nucleus. Because integration is a linear operation, adding two sources, even if they divide at different times, is trivial - we simply carry out the integration in equation~(\ref{DEsol}) for each nucleus that has already divided with the time properly offset, then sum over these nuclei. The only problem is to determine when each nucleus is supposed to divide. To find out, we perform what is essentially a numerical integration over time. We start with a release of material at the origin, which we model by having a delta function concentration there at $t=0$. By construction, the concentration field is then given by $G(\vec x, 0, t)$ as long as no other sources have released their chemicals into the system.  We proceed in small timesteps~$\Delta t$, calculating for each timestep the concentration at the location of each of the nuclei that have not yet divided, given the total concentration field generated by the nuclei that have divided so far. Suppose there is a total number of $N$ nuclei, $M$ of which have already released their chemicals. The $i$th source is located at $\vec x = \vec a_i$, released its chemicals at $t=t_i$, and the $t_i$'s are ordered. For $t_M < t < t_{M+1}$ we then find by using the superposition principle:
\begin{equation}
\label{csuperpos}
c(\vec x, t) = \sum_{i=1}^M \frac{Q}{4 \pi D (t-t_i)} \exp\left(-\frac{|\vec x - \vec a_i|^2}{4 D t}\right),
\end{equation}
where we have taken the number of dimensions to be two, and $Q$ is the number of chemicals released by a single nucleus. From $c(\vec x, t)$ we can determine when the next source will release its chemicals, by solving $c(\vec a_j, t) = \alpha$ for $M+1 \leq j \leq N$. We thus check all nuclei that have not yet released anything, and determine which one will be the next source by finding the one with the smallest value of $t$, which sets $t_{M+1}$. If there is a nonzero delay time between activation and release, we simply add it to the found value of $t_{M+1}$. Given the positions of the sources, the system thus has three parameters: the diffusion constant~$D$, the release concentration~$\alpha$, and the delay time~$t_\mathrm{delay}$. The `release-wave' is then the time at which a given source releases its chemicals versus its distance to the origin (i.e., the first source). An example is given in figure~3a of the main text, which reveals a clear wavefront with a well-defined wave speed. 

\subsubsection{Algorithm for finding wavefronts}
In summary, we use the following algorithm to numerically find the wavefronts within the biochemical diffusion model (and, with the proper adaptations, for the stress diffusion model as well):
\begin{itemize}
\item Generate a grid of hexagonally arranged nuclei with some small positional noise, centered at the origin.
\item Start with a delta function concentration at the origin at $t=0$.
\item Increase time in steps of $\Delta t$. For each timestep, calculate the local concentration at each of the sites of the nuclei, due to the nuclei that have released chemical so far. If one of these exceeds the critical concentration~$\alpha$, add a delta function concentration peak at this location and time.
\item Stop after either all nuclei have divided or a predefined time has been reached.
\end{itemize}

\subsubsection{Analysis of the steady-state wavefront}
In the case without delay time, it is fairly easy to determine the speed of the wavefront for the regime in which the wavefront is well-established, i.e., when its curvature is small. Suppose the lattice spacing is~$a$, the time it takes to get from one row to the next is~$t$, and the amount of material released by each nucleus is $Q$. The speed, in a triangular lattice, is then $v=a\cdot\frac12\sqrt{3} / t$, because the spacing between the two rows is $a\cdot\frac12\sqrt{3}$. To find the time, it turns out to be sufficient to only consider the 2 nearest neighbors in the previous row. We choose coordinates such that the neighbors are located at $(\pm \frac{a}{2}, 0)$, and our next nucleus is at $(0, \frac{a}{2}\sqrt{3})$. Then the time at which this next nucleus is activated is given by the solution of
\begin{equation}
\label{implicitt1}
\alpha = \frac{2Q}{4 \pi D t} \exp\left[ - \frac{a^2}{4 D t} \right].
\end{equation}
The results obtained using equation~(\ref{implicitt1}) are almost identical to those obtained by numerical solution of the full equations. If necessary, corrections could of course be made by including additional sources.

\begin{figure}[htb]
\begin{center}
\includegraphics{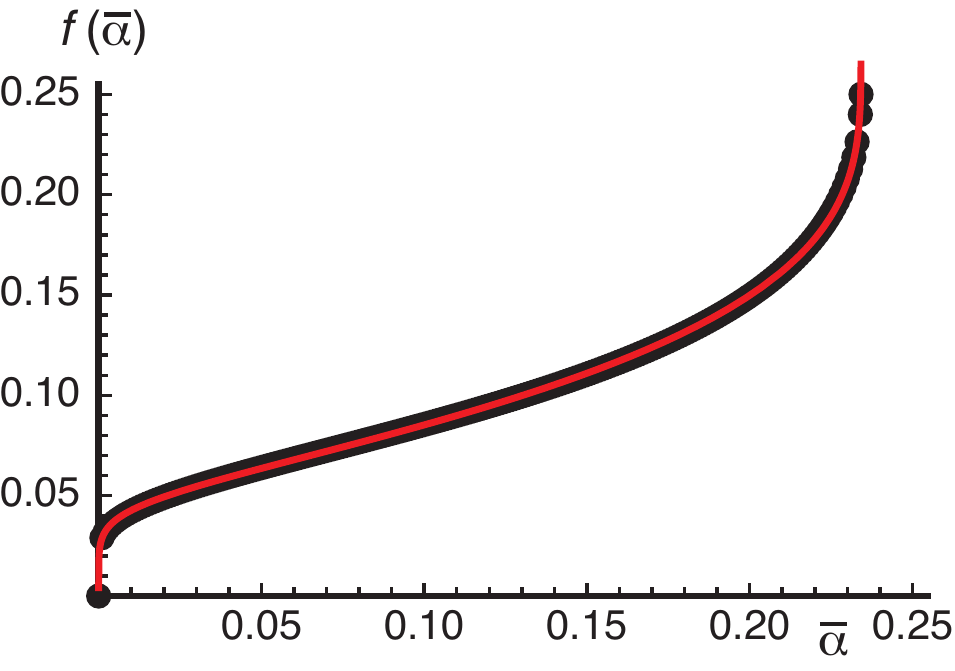}
\end{center}
\caption{Plot of~$f(\bar \alpha)$, defined in equation~(\ref{speed1}), determined numerically (black dots) and fitted with equation~(\ref{fapprox}) (red line).}
\label{fig:taualphanodelay}
\end{figure}

\begin{figure}[htb]
\begin{center}
\includegraphics{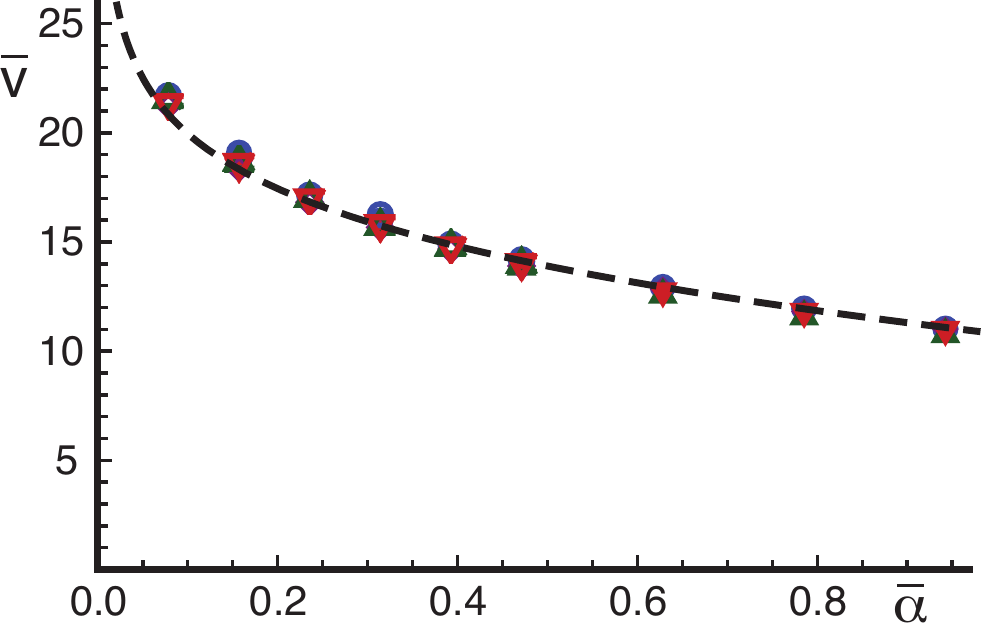}
\end{center}
\caption{Dimensionless wavefront speed~$\bar v = v/(D/a)$ as a function of the dimensionless threshold~$\bar \alpha = a^2 \alpha/Q$ in the biochemical signaling model. The figure shows $\bar v$ for several choices of the parameters $D$ and $a$ (different symbols). The data all collapse on a single curve, as described by equation~(\ref{speed1}). The dashed line is the fit to that curve of equation~(\ref{fapprox}).}
\label{fig:ffunctionfit}
\end{figure}

To analyze equation~(\ref{implicitt1}) further, we introduce dimensionless variables $\bar \alpha = a^2 \alpha/Q$ and $\tau = D t / a^2 $. The equation then becomes
\begin{equation}
\label{implicitt1b}
\bar \alpha = \frac{1}{2 \pi \tau} e^{-1/4\tau},
\end{equation}
which can of course not be solved analytically, but is easy to solve numerically. We note that the right hand side of~(\ref{implicitt1b}) has a maximum value of $\bar{\alpha}_{\max} = 2/\pi e$ at $\tau=1/4$, giving the condition that there can only be a wave if $\alpha < 2 Q / \pi e a^2$. If we write the inverse of~(\ref{implicitt1b}) as $\tau = f(\bar \alpha)$, we can write for the wavefront speed
\begin{equation}
\label{speed1}
v = \frac{D}{a} \frac12 \sqrt{3} \frac{1}{f(\bar \alpha)},
\end{equation}
so $v \propto D$ if both $a$ and $\alpha$ are fixed, but unfortunately the scaling of the speed with $a$ and $\alpha$ is hidden in $f(\bar \alpha)$. Based on the numerical determination of $f(\bar \alpha)$ we can capture its features fairly accurately with the following function
\begin{equation}
\label{fapprox}
f(\bar \alpha) \approx b_1 \bar \alpha^{1/n} \left[\frac14 - b_2 \left( \frac{8}{e}-\bar \alpha\right)^{1/m} \right],
\end{equation}
where fitting gives $b_1=1.38$, $b_2=0.28$, $n=6.3$ and $m=3.6$ (see figure~\ref{fig:taualphanodelay}). Our numerical solutions of the full equations show that the data do indeed collapse onto the curve described by equations~(\ref{speed1}) and~(\ref{fapprox}) for different diffusion coefficients and nuclear spacings (figure~\ref{fig:ffunctionfit}). In particular, we find that the wavefront speed $v$ always \textit{increases} as the nuclear spacing~$a$ \textit{decreases}, so $v$ increases with increasing cycle number.

\subsection{Wavefronts in the biochemical diffusion model with delay time}
\begin{figure}[htb]
\begin{center}
\includegraphics[scale=1]{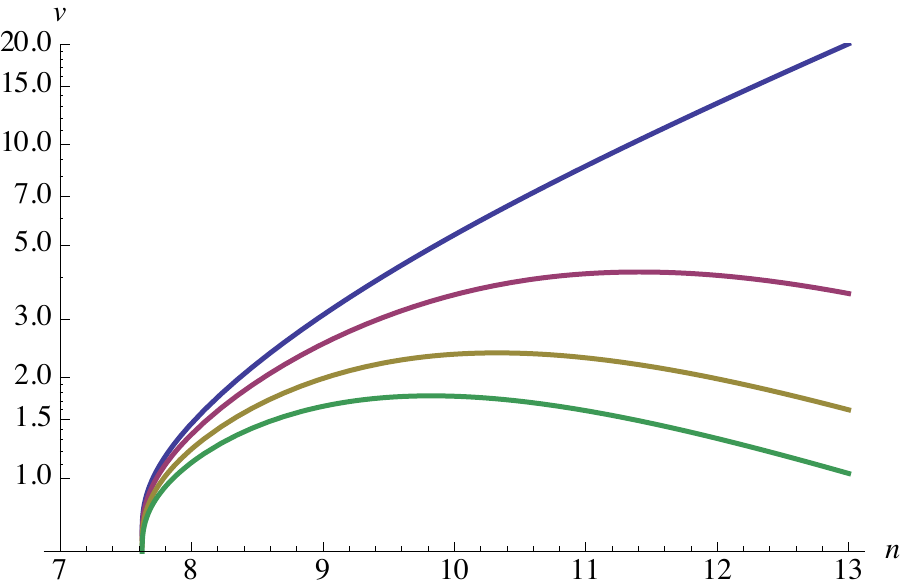}
\includegraphics[scale=1]{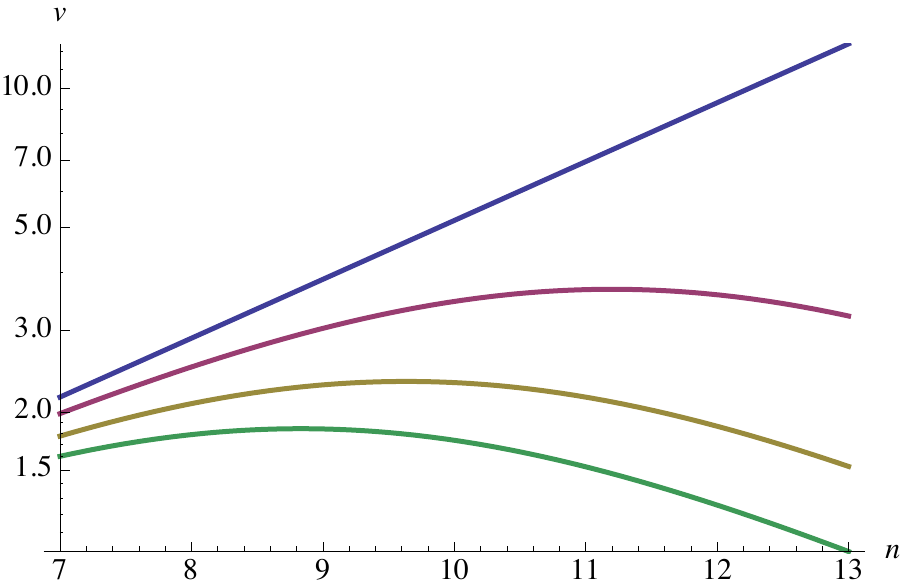}
\end{center}
\caption{Log-linear plots of the wavefront speed as a function of cycle number in the diffusion model with time delay. Top: fixed threshold $\alpha/Q = 0.001 \mu \mathrm{m}^{-2}$, $D = 10 \mu\mathrm{m}^2/\mathrm{s}$, and four values of $t_\mathrm{delay}$: 0s (blue), 2s (red), 5s (gold), and 8s (green). Bottom: same graph for fixed rescaled threshold $\bar{\alpha} =0.06$.}
\label{fig:diffmodelwithdelay}
\end{figure}

As indicated in Section~\ref{sec:biochemwavefronts}, we can include a delay time~$t_\mathrm{delay}$ into our diffusion model. Now a nucleus divides (i.e., releases its chemicals) a time $t_\mathrm{delay}$ \textit{after} the local concentration first reaches the threshold value~$\alpha$. To first approximation, the total time between the activation of a nucleus in a given row and one in the next row is simply the sum of the delay time and the travel time, which we found in the previous section. We therefore find:
\begin{equation}
\label{diffmodelspeedwithdelay}
v = \frac{D a \sqrt{3}}{2 D t_\mathrm{delay} + 2 a^2 f(a^2 \alpha/Q)}.
\end{equation}
Equation~\ref{diffmodelspeedwithdelay} breaks down for long delay times, or small thresholds, as in those cases the effects of earlier releases become important. However, for our system these effects are small. Figure~\ref{fig:diffmodelwithdelay} shows the wavefront speed~$v$ as a function of cycle number for four different values of the delay time, and for both the cases that $\alpha$ and $\bar{\alpha}$ are constant. Note that for constant~$\alpha$ there is a lower limit to the cycle number below which the model predicts no wavefronts (as the activation threshold is not reached). Note also that although the introduction of the delay time causes the increasing trend in the wavefront speed to chance in the later cycles, it will still increase in earlier cycles.

\subsubsection{Fitting the data with time delay}
Naturally, the more parameters we have, the easier it is to fit any set of experimental data points. In the diffusion model with delay time we have four parameters: the diffusion constant~$D$, the grid size~$a$, the threshold value~$\alpha$ and the delay time~$t_\mathrm{delay}$. The grid size (i.e. the spacing between the nuclei) is measured independently, leaving us with three parameters which we can vary. Reasonable values for the diffusion constant for a small chemical in the early \textit{Drosophila} embryo are $D=5-100$~$ \mu\mathrm{m}^2/\mathrm{s}$, as measured by Gregor et al.~\cite{gregor05}. As indicated in the main text, we cannot fit the experimental trend (a wavefront speed that decreases exponentially with cycle number) with fixed values for $t_\mathrm{delay}$ and $\alpha$ for any value of $D$ within this range (see figure~3b of the main document). The only way we can thus fit the experimental data within this model is if either (or both) of $\alpha$ and $t_\mathrm{delay}$ change with the cycle number.

We have systematically investigated a number of options, changing $\alpha$ or $t_\mathrm{delay}$ with cycle number. Some results are given in table~\ref{table:diffmodelfits}.  We did not find any result that fit the data in which the numbers change in a well-defined way (e.g. the delay time increasing linearly or exponentially with the cycle number). Moreover, in the case of variable delay time, we find that in the first cycle (cycle 10), the interaction is between nearest-neighbors as in the model without time delay, but the interaction range goes up every cycle, up to 5 rows apart in cycle 13. In the case of variable concentration threshold, we need a change of at least an order of magnitude in each cycle to fit the experimental data. Even in the case where we allow both variables to change, we keep finding at least one of these two problems. Even though we cannot strictly rule out the diffusion model with variable time delay and concentration threshold, these results make it very unlikely that this model is actually correct.

\begin{table}[hp]
\begin{center}
\begin{tabular}{|c|c|c|c|c|c|}
\hline
cycle & nuclear spacing & wavefront speed & $t_\mathrm{delay}$ (s) & $\alpha/Q$ ($10^{-4}/\mu\mathrm{m}^2$) & ($t_\mathrm{delay}$, $\alpha/Q$) \\
number & ($\mu\mathrm{m}$) & ($\mu\mathrm{m}/\mathrm{s}$) & ($\alpha$ fixed) & ($t_\mathrm{delay}$ fixed) & (s, $10^{-4}/\mu\mathrm{m}^2$) \\
\hline
10 & 23.4 $\pm$ 0.8 & 2.9 $\pm$ 0.9 & 1.5 & 0.0044 & (5, 0.254) \\
11 & 18.2 $\pm$ 0.6 & 2.2 $\pm$ 0.9 & 5.7 & 0.29 & (10, 0.300) \\
12 & 13.2 $\pm$ 0.3 & 1.5 $\pm$ 0.4 & 16.0 & 6.9 & (20, 0.685)\\
13 & 9.7 $\pm$ 0.2 & 1.0 $\pm$ 0.2 & 36.1 & 31 & (40, 1.057) \\
\hline
\end{tabular}
\end{center}
\caption{Experimentally determined values of the nuclear spacing and wavefront speed (set 1), and numerically determined values of the required delay time $t_\mathrm{delay}$ and threshold concentration $\alpha$ to fit the experimental data. In column four, $D=15$~$\mu\mathrm{m}^2/\mathrm{s}$ and $\alpha/Q = 5 \cdot 10^{-4}$~$\mu\mathrm{m}^{-2}$; in column 5,  $D=10$~$\mu\mathrm{m}^2/\mathrm{s}$ and $t_\mathrm{delay} = 10$~$\mathrm{s}$; in column 6, $D=10$~$\mu\mathrm{m}^2/\mathrm{s}$ and $t_\mathrm{delay}$ is assumed to double in each cycle. None of the columns show a systematic dependency of the parameters on the cycle number, making it impossible to assign predictive power to the numbers found, or to find a model to explain the dependencies. Note also that for the case of fixed delay time (column 5), we need to assume that the threshold value goes up by at least an order of magnitude in each cycle. In the case of variable delay time (columns 4 and 6), the interactions become very long-ranged in the later cycles, with nuclei up to 5 rows apart triggering each other in cycle 13, even though in cycle 10 the interaction only involves nearest neighbors.}
\label{table:diffmodelfits}
\end{table}

\subsection{Wavefronts in the mechanical model}
The analysis leading to wavefront propagation in the mechanical model is described in Ref.~\cite{idema13prl}. As in the case of diffusion, we must set a threshold to determine when a source (a nucleus) is activated. The simplest option is to look at the eigenvalues of the stress tensor: if the largest of those (taking absolute values) exceeds a certain threshold~$\alpha$, the nucleus is activated. An activated nucleus adds an additional force dipole term to the stress field in the system, which in turn of course affects the displacement field, as described by the Green tensor solution of equation~(\ref{mechanicalmodel}). Note that we assume linear elasticity, so that the strain is linear in the displacement, and the stress is linear in the strain. The superposition principle therefore holds not just for the displacements but for the strain and stress fields as well. 

\begin{figure}[htb]
\begin{center}
\begin{tabular}{c}
\includegraphics{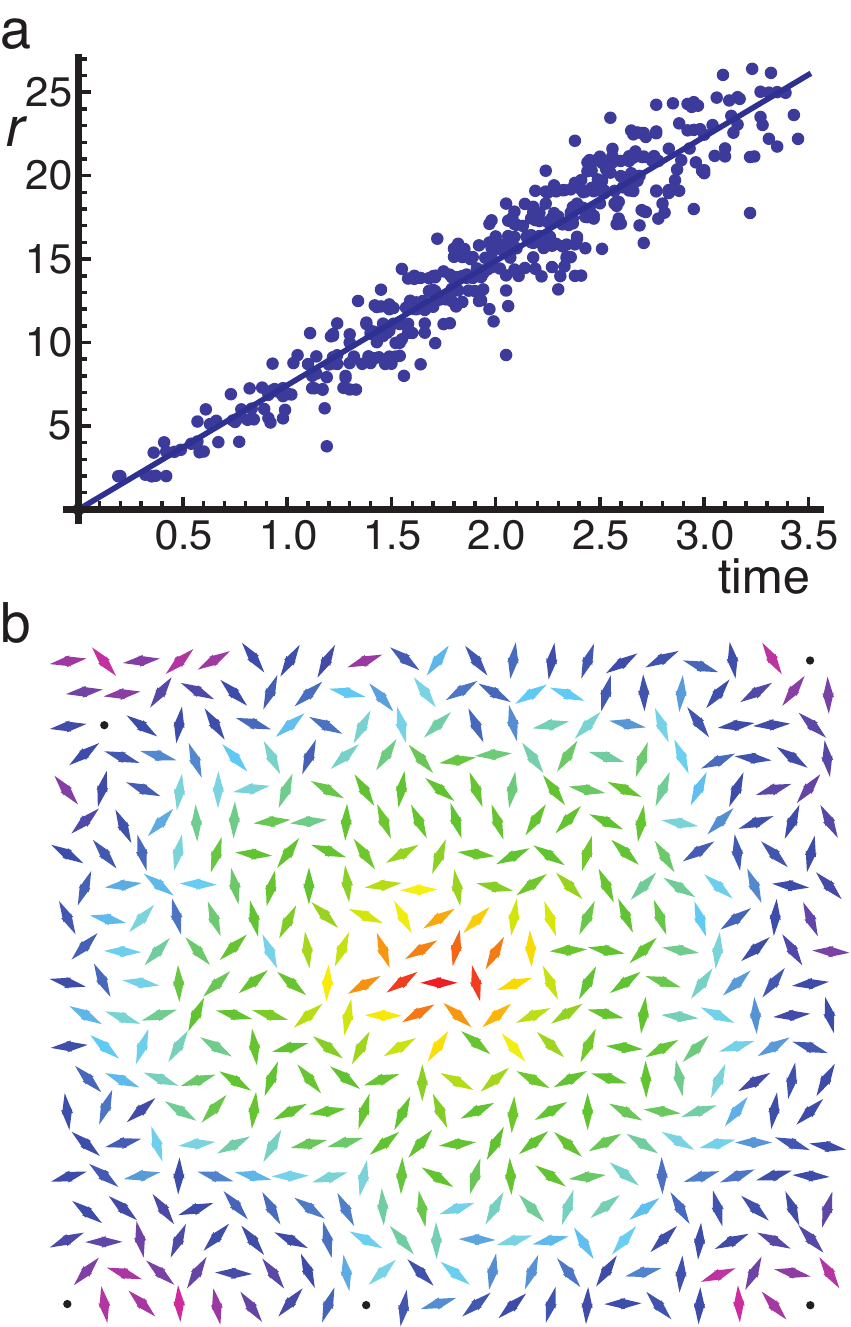}
\end{tabular}
\end{center}
\caption{Wavefront from a simulation with $21 \times 21$ nuclei. The nuclei are arranged on a hexagonal grid with random small offsets. The wave starts at the center point which generates a stress dipole of unit strength along the $x$-axis at $t=0$. Whenever the absolute value of the largest eigenvalue of the stress tensor at another nucleus exceeds the threshold value $\alpha$, it also divides, adding a unit stress dipole in a random direction to the total stress field. a) Distance of the dividing nuclei to the center vs. their activation time, with a linear fit. b) Graphical representation of the 2 dimensional field, with the dipoles indicated in the orientation in which they divide, and color-coded according to the time that they divide, on a hue scale (red-yellow-green-blue-violet). 
}
\label{fig:wavefront2}
\end{figure}

The implementation of the stress-mediated signaling model follows the same pattern as that of the chemical-diffusion-mediated signaling model, with the concentration~$c$ replaced by the stress tensor~$\sigma_{ij}$. An example implementation on a grid of $21 \times 21$ nuclei is shown in figure~\ref{fig:wavefront2}. The figure shows a clear wavefront which has a well-defined speed.

As in the diffusion model, we analyze our mechanical model in terms of dimensionless parameters. There is only one quantity in our model that has the dimensions of a speed, namely $\mu / a \Gamma$, which means that the resulting wavefront speed has to scale linearly with this factor, as indeed it does. We again define a dimensionless wavefront speed $\bar{v} = \frac{a \Gamma}{\mu} v = \frac{v}{D/a}$ and a dimensionless stress threshold $\bar{\alpha} = a^2 \alpha/Q$, where $Q$ is now the strength of the force dipole. We can then write
\begin{equation}
\label{mechanicalwavefrontspeed}
v = \frac{\mu}{a \Gamma} g(\bar{\alpha}, \nu)
\end{equation}
We determine the function $g(\bar{\alpha}, \nu)$ by numerically solving the model. As detailed in~\cite{idema13prl}, we find that it can be well described by the following functional form, derived using a similar argument we used to arrive at equation~(\ref{implicitt1b}) for the diffusion model:
\begin{equation}
\label{gfunction}
g(\bar{\alpha}, \nu) = -4\frac{(c_1+c_2\bar{\alpha}) \log(\bar{\alpha})}{1-\nu^2} + c_3.
\end{equation}
Note that the form in equation~(\ref{gfunction}) differs slightly from that in~\cite{idema13prl} because of the use of the shear modulus~$\mu$ instead of the Young's modulus~$E$ in the definition of $\bar{v}$, and the extra constant~$c_3$, which is due to the introduction of semi-periodic boundary conditions. We use equations~(\ref{mechanicalwavefrontspeed}) and~(\ref{gfunction}) with $c_1=4.5$, $c_2=1.5$ and $c_3=5.4$ to fit the experimental data in figure 3f of the main text.

\newpage

\end{document}